\documentclass[12pt,prd,superscriptaddress,nofootinbib,tightenlines,preprintnumbers]{revtex4}

\usepackage{amstext,amssymb}
\usepackage{amsmath}
\usepackage{graphicx}
\usepackage[hyperfootnotes=false]{hyperref}
\usepackage{xspace}
\usepackage{color}
\usepackage{units}
\usepackage{slashed} 

\def \L {\mathcal{L}} 

\def \vec#1{{\boldsymbol{#1}}}
\newcommand{\hc}{\ensuremath{\text{h.c.}}}

\begin{document}

\title{Common Origin of $3.55$~keV X-ray line and Gauge Coupling Unification with Left-Right Dark Matter}
\author{Debasish Borah}
\email{dborah@iitg.ernet.in}
\affiliation{Department of Physics, Indian Institute of Technology Guwahati, Assam 781039, India}

\author{Arnab Dasgupta}
\email{arnabdasgupta28@gmail.com}
\affiliation{Theoretical Physics Division, Physical Research Laboratory, Navrangpura, Ahmedabad, Gujarat 380009, India}
\affiliation{Institute of Physics, Sachivalaya Marg, Bhubaneswar 751005, India}

\author{Sudhanwa Patra}
\email{sudhanwa@iitbhilai.ac.in}
\affiliation{Department of Physics, Indian Institute of Technology Bhilai, Govt. Engg. College Campus, Sejbahar, Raipur, Chhattisgarh 492015, India}
\affiliation{Center of Excellence in Theoretical and Mathematical Sciences,
Siksha `O' Anusandhan University, Bhubaneswar 751030, India}

\begin{abstract}

We present a minimal left-right dark matter framework that can simultaneously explain the recently observed 3.55 keV X-ray line from several galaxy clusters and gauge coupling unification at high energy scale. Adopting a minimal dark matter strategy, we consider both left and right handed triplet fermionic dark matter candidates which are stable by virtue of a remnant $$\mathcal{Z}_2\simeq (-1)^{B-L}$$ 
symmetry arising after the spontaneous symmetry breaking of left-right gauge symmetry to that of the standard model. A scalar bitriplet field is incorporated whose first role is to allow radiative decay of right handed triplet dark matter into the left handed one and a photon with energy 3.55 keV. The other role this bitriplet field at TeV scale plays is to assist in achieving gauge coupling unification at a high energy scale within a non-supersymmetric $SO(10)$ model while keeping the scale of left-right gauge symmetry around the TeV corner. Apart from solving the neutrino mass problem and giving verifiable new contributions to neutrinoless double beta decay and charged lepton flavour violation, the model with TeV scale gauge bosons can also give rise to interesting collider signatures like diboson excess, dilepton plus two jets excess reported recently in the large hadron collider data.
\end{abstract}
%

\maketitle


\section{Introduction}
\label{sec:introduction}
Since the observations of galaxy rotation curves made by Fritz Zwicky \cite{zwicky}, the evidence suggesting the presence of dark matter in the Universe has been ever increasing with the latest cosmology experiment Planck suggesting around $26\%$ of the present Universe's energy density being made up of dark matter \cite{Planck13}. In terms of density parameter $\Omega$, the dark matter abundance in the present Universe is reported as
\begin{equation}
\Omega_{\text{DM}} h^2 = 0.1187 \pm 0.0017
\label{dm_relic}
\end{equation}
where $h = \text{(Hubble Parameter)}/(100 \text{km} \text{s}^{-1} \text{Mpc}^{-1})$ is a parameter of order unity. In spite of growing astrophysics and cosmological evidence in support of dark matter, its particle nature is yet unknown. The characteristics a particle dark matter candidate should satisfy \cite{bertone} rule out all the particles in the standard model (SM) of particle physics as dark matter candidates. With the dark matter direct detection experiments \cite{Aprile:2013doa, LUX, PandaXII, Xenon1T} giving null results as of now, there have been growing efforts in the indirect detection frontiers as well. Recently, one promising indirect signature of dark matter was reported by two independent analysis \cite{Xray1} and \cite{Xray2} of the data collected by the XMM-Newton X-ray telescope. Their analysis hinted towards the existence of a monochromatic X-ray line with energy 3.55 keV in the spectrum of 73 galaxy clusters. The analysis \cite{Xray1} also claimed the presence of the same line in the Chandra observations of the Perseus cluster. Later on, the same line was also found in the Milky Way by analysing the XMM-Newton data \cite{Xray3}. Although the analysis of the preliminary data collected by the Hitomi satellite (before its unfortunate crash) do not confirm such a monochromatic line \cite{hitomi1}, one still needs to wait for a more sensitive observation with future experiments to have a final word on it. Interestingly, the authors of \cite{hitomi2} considered a specific dark matter model to show consistency among Hitomi, XMM-Newton and Chandra observations. More recently, the authors of \cite{xray17} have reported a $3\sigma$ detection of a $3.5$ keV emission line in the spectrum of the Cosmic X-ray background using Chandra observations towards the COSMOS Legacy and CDFS survey fields.

If there is no astrophysical source behind the origin of this line, then it is very tempting to study a possible dark matter origin of this line. As pointed out by the two teams analysing the above-mentioned data, such a monochromatic line can be naturally incorporated within the framework of sterile neutrino dark matter where a sterile neutrino with 7.1 keV mass decays into a photon and a SM neutrino. Different particle physics models with keV scale sterile neutrino dark matter as possible explanation of the 3.55 keV X-ray line were studied in \cite{Xraysterile}. A few other theoretical models were also suggested in \cite{Xrayothers1, Xrayothers2}. Although it is more natural to connect the keV X-ray line with a dark matter particle having similar mass, it is nevertheless worth exploring other possible scenarios. For example, the origin of this 3.55 keV line from electroweak scale dark matter candidates also found some attention in the works \cite{Xrayweakscale, Falkowski:2014sma}. Typically the keV sterile neutrino gives rise to a scenario called warm dark matter (WDM) whereas electroweak scale dark matter is a popular cold dark matter (CDM) candidate and they, in general have very different consequences in astrophysical structure formations. Although it involves fine-tuning in various couplings in a model connecting CDM and keV line, it also has some advantage compared to WDM scenario. For example, according the analysis \cite{Xray1, Xray2}, a keV scale sterile neutrino should have mixing with the SM neutrinos of the order $\approx 10^{-11}-10^{-10}$ to explain the observed X-ray line. Since the sterile neutrinos interact with the known SM particles only through this mixing, such a tiny value of mixing angle never allows the sterile neutrino dark matter to enter thermal equilibrium in the early Universe. The puzzle related to the production of sterile neutrino dark matter in the Universe therefore, invites additional new physics. However, CDM is well understood due to their standard weak interaction cross sections falling under the regime of weakly interacting massive particle (WIMP), the most widely studied dark matter framework in the literature.

Although there have been a huge number of WIMP or CDM models in the particle physics literature, it is more interesting to study those models which not only predicts a stable CDM candidate but also provides answers to other open questions in particle physics. The left-right symmetric model (LRSM) \cite{lrsm,lrsmpot} is one such highly motivated framework studied extensively in the last few decades from several beyond standard model (BSM) motivations. The model not only explains the origin of parity violation in weak interactions but also explains the origin of tiny neutrino masses naturally. The symmetry group of the LRSM can also be embedded within grand unified theory (GUT) symmetry groups like $SO(10)$ providing a non-supersymmetric framework to achieve gauge coupling unification. Recently, this model received lots of attention in view of the hints of large hadron collider (LHC) about the existence of new physics around the TeV corner: CMS $eejj$ excess \cite{CMSeejj}, ATLAS diboson excess \cite{diboson} and very recently, the 750 GeV diphoton excess \cite{atlasconf,lhcrun2a,CMS:2015dxe}. Although the subsequent updates from the LHC experiments \cite{LHC16update} did not confirm their preliminary results hinting the existence of this 750 GeV resonance, the usual motivations to study LRSM have remained the same. From dark matter phenomenology point of view also this model was recently studied \cite{Heeck:2015qra,Garcia-Cely:2015quu} in a spirit of minimal dark matter framework \cite{Cirelli:2005uq,Garcia-Cely:2015dda,Cirelli:2015bda}. In this framework, new particles with high $SU(2)$ dimensions included in the minimal LRSM are either accidentally stable due to the absence of renormalizable couplings that can allow their decay into SM particles or due to a remnant discrete symmetry which protects the dark matter from decaying. In this work, we explore this possibility further with a goal of explaining the 3.55 keV X-ray line from dark matter in LRSM. Although it is possible to have sterile neutrino dark matter in LRSM, it is difficult to generate the correct relic abundance of a 7.1 keV sterile neutrino dark matter if the right handed gauge boson mass is kept around 3 TeV \cite{wdmlr}. Therefore, here we pursue the scenario of CDM origin of the 3.55 keV line \cite{Xrayweakscale, Falkowski:2014sma} by considering a two component left-right dark matter model with a keV mass splitting between the dark matter candidates. To be more specific, we incorporate both left and right handed fermionic triplet into minimal LRSM such that the neutral components of them can be stable CDM candidates. Although the left-right symmetry predicts their masses to be degenerate in the minimal LRSM, we consider a tiny mass splitting between the two dark matter candidates. Such a tiny splitting can occur naturally by the vacuum expectation value (vev) of the neutral component of a scalar bitriplet field or due to the parity breaking effects at high energy scale. We show that only the latter way of generating the mass splitting is acceptable from dark matter phenomenology. We also show how the bitriplet with TeV scale mass can assist in heavier dark matter decay into the lighter one at two loop level with a lifetime required to explain the X-ray data. Interestingly, the same bitriplet with mass around the TeV corner assists in achieving gauge coupling unification at a high scale by keeping the scale of left-right symmetry within the reach of LHC.

This letter is organised as follows. In section \ref{sec1}, we briefly discuss the model. In section \ref{sec2} we discuss the possibility of generating the 3.55 keV line within our framework followed by an outline of the relic abundance calculation of dark matter candidates in section \ref{sec3}. In section \ref{sec4} we show the possibility of gauge coupling unification in the model and finally conclude in section \ref{sec6}.

\section{The Model Framework}
\label{sec1}
Left-right symmetric model is one of the most widely studied BSM frameworks in last few decades due to the natural origin 
of tiny neutrino mass and spontaneous parity violation. The symmetry group of the standard model is extended by an additional $SU(2)_R$ group under 
which the right handed fermions transform as doublets similar to the way left handed fermions transform under $SU(2)_L$. The $U(1)_Y$ of SM is replaced 
with an anomaly free $U(1)_{B-L}$ providing a better understanding of electromagnetic charges of fundamental particles after electroweak symmetry breaking. 
The model also has an in built discrete symmetry $Z_2$ which interchanges left and right handed fields thereby making the couplings in both the sectors equal 
(related) or left-right symmetric. Under the symmetry group $SU(2)_L \times SU(2)_R \times U(1)_{B-L} 
\times SU(3)_C$ of LRSM, the usual quarks and leptons transform as
\begin{eqnarray}
& &q_{L}=\begin{pmatrix}u_{L}\\
d_{L}\end{pmatrix}\equiv[2,1,{\frac{1}{3}},3], \quad q_{R}=\begin{pmatrix}u_{R}\\
d_{R}\end{pmatrix}\equiv[1,2,{\frac{1}{3}},3]\,,\nonumber \\
& &\ell_{L}=\begin{pmatrix}\nu_{L}\\
e_{L}\end{pmatrix}\equiv[2,1,-1,1], \quad \ell_{R}=\begin{pmatrix}\nu_{R}\\
e_{R}\end{pmatrix}\equiv[1,2,-1,1] \,, \nonumber 
\end{eqnarray}
while the symmetry breaking is implemented with the following scalar bidoublet $\Phi$ as well as triplet scalars $\Delta_{L,R}$ 
with their matrix representations, 
$\Delta_{L,R}$ as
\begin{eqnarray}
&&\Phi \equiv \begin{pmatrix} \phi_1^0 & \phi_2^+ \\ \phi_1^- & \phi_2^0 \end{pmatrix}\equiv[2,2,0,1] \, , \quad \nonumber \\
&&\Delta_{L} \equiv \begin{pmatrix} \delta_{L}^+/\sqrt{2} & \delta_{L}^{++} \\ \delta_{L}^0 & -\delta_{L}^+/\sqrt{2} \end{pmatrix} \equiv[3,1,2,1] \, , \quad 
 \Delta_{R} \equiv \begin{pmatrix} \delta_{R}^+/\sqrt{2} & \delta_{R}^{++} \\ \delta_{R}^0 & -\delta_{R}^+/\sqrt{2} \end{pmatrix} \equiv[1,3,2,1] \, .
\nonumber 
\end{eqnarray}
The spontaneous symmetry breaking of the left-right symmetric gauge group to the SM is achieved by assigning a non-zero vev to the neutral component of the right-handed scalar triplet $\Delta_R$ while the Higgs doublets $\phi$ contained in the scalar bidoublet $\Phi$ break the SM gauge group to the low energy $U(1)_Q \times SU(3)_C$ symmetry as follows
$$ SU(2)_L \times SU(2)_R \times U(1)_{B-L} \times SU(3)_C \quad \underrightarrow{\langle
\Delta_R \rangle} \quad SU(2)_L\times U(1)_Y \times SU(3)_C \quad \underrightarrow{\langle \phi \rangle} \quad U(1)_{Q} \times SU(3)_C$$
After the spontaneous symmetry breaking $SU(2)_L \times SU(2)_R \times U(1)_{B-L} 
\times SU(3)_C \rightarrow \text{SM} \rightarrow U(1)_Q \times SU(3)_C$, the electromagnetic charge of the components of above fields arise as
\begin{align}
Q=T_{3L}+T_{3R}+\frac{B-L}{2} 
\end{align}
The scalar triplets not only play the role of breaking the initial gauge symmetry into that of the SM spontaneously 
but also generate Majorana masses of heavy and light neutrinos. The scalar bidoublet is needed for spontaneous 
breaking of SM gauge symmetry to low energy theory of $U(1)_Q \times SU(3)_C$ so as to give correct masses 
to electroweak vector bosons and SM charged fermions. 

\section{Explanation for $3.55$~keV X-ray line} 
\label{sec2}
We introduce a pair of fermion triplets $\Sigma_{L,R}$ with the following matrix representation 
\begin{eqnarray}
&&\Sigma_{L}=\begin{pmatrix}
  \Sigma^0_{L}  & \sqrt{2} \Sigma^+_{L}  \\
  \sqrt{2} \Sigma^-_{L} & -\Sigma^0_{L}
 \end{pmatrix} \equiv [3,1,0,1] \, , \nonumber \\
&&\Sigma_{R}=\begin{pmatrix}
  \Sigma^0_{R}  & \sqrt{2} \Sigma^+_{R}  \\
  \sqrt{2} \Sigma^-_{R} & -\Sigma^0_{R}
 \end{pmatrix} \equiv [1,3,0,1] \, .
\end{eqnarray}
where the neutral component of the each fermion triplet can be a stable dark matter candidate. The stability of dark matter multiplets in LRSM is either ensured automatically 
because of high $SU(2)$~dimension which forbids the tree level decay or due to a remnant $$\mathcal{Z}_2\simeq (-1)^{B-L}$$ 
symmetry arising after the spontaneous symmetry breaking of LRSM down to SM gauge group i.e, $SU(2)_R \times U(1)_{B-L} \to U(1)_Y$. Under this 
remnant discrete symmetry $\mathcal{Z}_2\simeq (-1)^{B-L}$, the usual leptons are odd while all bosons including scalars and gauge bosons are even. Since fermion triplets have vanishing $B-L$ charge, they are even under the remnant discrete symmetry, prohibiting them from decaying into SM leptons which are $\mathcal{Z}_2$ odd. For detailed discussion, one may refer to \cite{Heeck:2015qra,Garcia-Cely:2015quu}. Due to the chosen transformation of fermion triplets under the gauge symmetry of LRSM, one can write down their bare mass terms 
$M_{\Sigma} \Sigma_{L, R} \Sigma_{L, R}$ in the Lagrangian where $M_{\Sigma}$ is same for both $\Sigma_L, \Sigma_R$ due to the in built left-right discrete symmetry. As mentioned earlier, 
our strategy to generate the 3.55 keV X-ray line from left-right dark matter is to induce a mass splitting of 3.55 keV between the neutral components 
of fermion triplets $\Sigma^0_{L,R}$. Although this can not be done explicitly due to the in built left-right symmetry, one can achieve it in two different ways:\\

$(a)$ \textit{By giving a vev to the neutral component of the bitriplet scalar field}: The bitriplet field introduced to allow radiative decay of heavier dark matter into the lighter dark matter and a photon can also induce a tiny mass splitting between two dark matter candidates, which were degenerate in masses due to the in built left-right symmetry. This scalar bitriplet field $\psi \sim (\textbf{3},\textbf{3},0,1)$ can be written in matrix form as
\begin{equation}
\psi = \left(\begin{array}{ccc}
\zeta^{0*} & \epsilon^+ & \zeta^{++} \\
-\zeta^{+*} & \epsilon^0 & \zeta^{+} \\
\zeta^{++*} & -\epsilon^{+*} & \zeta^0 \\
\end{array}
\right)
\end{equation}
The neutral component of the above field can acquire a tiny induced vev after the electroweak symmetry breaking and can generate the required mass splitting. 
Thus, the relevant interaction Lagrangian is given by
\begin{align}
\mathcal{L}_\Sigma \supset M_\Sigma \left(\Sigma^T_L C \Sigma_L +\Sigma^T_R C \Sigma_R \right)
+ \frac{Y_{\psi}}{2} \overline{\Sigma_L} \psi \Sigma_R \,,
+\,  \text{h.c.}
\end{align}
After inclusion of Yukawa interaction between both left- and right-handed fermion triplets with scalar bitriplet, a small mass splitting will arise between the two neutral components $\Sigma^0_L$ and $\Sigma^0_R$. 
The mass matrix for the neutral fermion triplets in the basis $(\Sigma^0_L, \Sigma^0_R)$ after scalar bitriplet 
get its usual vev is given by
\begin{equation}
\begin{pmatrix} M_{\Sigma}  & \delta M_{\Sigma} \\
\delta M_{\Sigma}  & M_{\Sigma} \end{pmatrix}
\label{massmatrix}
\end{equation}
where $\delta M_{\Sigma}=\frac{Y_{\psi}}{2} \langle \epsilon^0 \rangle$. 
Diagonalising the above mass matrix gives the following physical masses
\begin{equation}
M_{1} = M_{\Sigma} - \frac{Y_{\psi}}{2} \langle \epsilon^0 \rangle \, , \quad 
M_{2} = M_{\Sigma} + \frac{Y_{\psi}}{2} \langle \epsilon^0 \rangle
\end{equation}
This results in a mass splitting between two mass eigenstates $\Sigma_1$ and $\Sigma_2$ as $\delta M \equiv M_{\Sigma_2}-M_{\Sigma_1} =2 \delta M_{\Sigma}$. 
Thus the heavier and lighter dark matter candidates have a mass difference of  $\delta M = Y_{\psi} \langle \epsilon^0 \rangle$ which is 3.55 keV in this case. 
Such a tiny mass splitting also kinematically forbids any tree level decay of the heavier  dark matter particle into the lighter one and standard model particles except perhaps, to neutrinos, as we discuss below. Although this is a minimal scenario with the same bitriplet being responsible for one loop decay of heavier dark matter as well as generating mass splitting between the dark matter candidates, it suffers from a serious problem. Since the mass matrix \eqref{massmatrix} introduces a large mixing $\theta \approx \pi/4$ between $\Sigma^0_L$ and $\Sigma^0_R$. This allows the heavier dark matter decay into the lighter one and a photon through a $W_L$ boson mediated process. For TeV scale fermion triplets, this decay width comes out to be around $10^{-12}, 10^{-15}$ GeV for heavier dark matter masses $1, 3$ TeV respectively. This decay width is far too larger than what is required to generate the $3.55$ keV line, as discussed below. \\

$(b)$ \textit{Through parity breaking effects at high scale}: Even if the neutral component of the bitriplet does not acquire any vev, and its sole role is to allow the decay of heavier dark matter into lighter dark matter and a photon at two loop level, one can generate such a tiny splitting due to parity breaking effects at high energy scale. As it will be discussed in section \ref{sec4}, it is preferable to break the discrete parity spontaneously at a high energy scale while keeping the gauge symmetry of LRSM unbroken down to the TeV scale. This can be done by introducing a parity odd singlet $\eta$ which couple to $\Sigma_L$ and $\Sigma_R$ with an opposite relative sign. If $\eta$ gets a non-zero vev, it will break the degeneracy in the masses of $\Sigma_L$ and $\Sigma_R$ as
$$M_2 = M_{\Sigma} + \lambda \langle \eta \rangle, \;\; M_1 = M_{\Sigma} - \lambda \langle \eta \rangle$$
This generates a mass splitting of $2\lambda \langle \eta \rangle$ between the two dark matter particles. Here $\lambda$ is the dimensionless coupling between $\eta$ and fermion triplets. From unification point of view, the parity breaking is desired to occur near the unification scale, close to $10^{16}$ GeV. However, such a large vev of $\eta$ will completely decouple the heavier dark matter candidate from the low energy phenomenology. To keep both the dark matter candidates near the TeV scale with a tiny mass difference of 3.55 keV, we assume the effect of parity breaking on the mass splitting to be very small. This is equivalent to taking very small values of $\lambda$. This can occur naturally, if $\eta$ is charged under some global symmetry, such that the term $ \lambda \eta \Sigma_{L,R} \Sigma_{L,R}$ breaks the symmetry explicitly. Since $\lambda \rightarrow 0$ results in an enhanced the global symmetry, its value can be naturally small. Since $\Sigma_L$ does not mix with $\Sigma_R$ in this case, the problem of fast decay through $W_L$ bosons does not occur in this case. We do our follow up calculation by considering this setup. However, such tiny breaking of global symmetry leads to a pseudo-Nambu-Goldstone boson whose mass is expected to be small since all the global symmetry breaking terms in the scalar potential are small by naturalness argument. Another way to generate a small $\lambda$ is to generate it effectively through higher dimensional operators. For example if $\Sigma_{L,R}$ is charged under a discrete symmetry $Z_N$ with a nontrivial charge $z$ (so tha $z^N=1$), then their masses can be generated by a singlet scalar $\zeta$ with $Z_N$ charge $z^{N-2}$. Although the tree level masses are degenerate $M_{\Sigma} = \lambda_{\zeta} \langle \zeta \rangle$, the parity odd singlet $\eta$ can break this degeneracy due to dimension five operator like $\lambda_{\zeta} \zeta \eta (\Sigma^T_R C \Sigma_R-\Sigma^T_L C \Sigma_L)/\Lambda$. This generates an effective $\lambda = \frac{\lambda_{\zeta} \langle \zeta \rangle}{\Lambda}$ that can be naturally small.

However, at low energies, the $SU(2)_L$ and $SU(2)_R$ gauge interactions can induce left-right symmetry breaking effects and since $M_W \neq M_{W_R}$, these effects will further induce a mass splitting between left and right fermion triplets. These effects, which can also arise if $g_L \neq g_R$, can generate this mass splitting at one-loop level through gauge interactions. As we discuss later, these are the same loop diagrams that also break the degeneracy between charged and neutral components of a fermion triplet. These one-loop corrections to the masses of the neutral components of left and right fermion triplets by gauge interactions are given respectively by
\begin{equation}
\triangle M^0_L = \frac{g^2_L}{16\pi^2}  (2M_{\Sigma}) \left[ -f(r_L)+\cdot \cdot \cdot\right]
\end{equation}
\begin{equation}
\triangle M^0_R = \frac{g^2_R}{16\pi^2}  (2M_{\Sigma}) \left[ -f(r_R)+\cdot \cdot \cdot\right]
\end{equation}
where $r_L = \frac{M_W}{M_{\Sigma}}, r_R = \frac{M_{W_R}}{M_{\Sigma}}$ with the loop function $f(r)$ being given as
\begin{equation}
f(r) \equiv 2 \int^1_0 dt (1+t) \ln\left[t^2+(1-t)r^2 \right]
\label{eq:fr}
\end{equation}
In the above expressions for one-loop mass corrections, the dots represent the common terms that contain a divergent term, the logarithmic term proportional to $\ln M_{\Sigma}$ etc. Although these individual one-loop corrections are divergent, their difference is finite and calculable and gives rise to a mass splitting between the neutral components of the left and right fermion triplets given by
\begin{equation}
M_{\Sigma^0_L}-M_{\Sigma^0_R} = \frac{1}{8\pi^2} \left[g^2_R f(r_R) -g^2_L f(r_L) \right] M_{\Sigma}
\end{equation}
For $g_L = g_R =g$, the electroweak gauge coupling and $M_{\Sigma} =3$ TeV, $M_{W_R}=5$ TeV, this splitting is rather large and turns out to be approximately $84$ GeV, which is much larger than the desired splitting of 3.55 keV. 

Therefore, we need a fine-tuning between the splittings through parity breaking scale at high scale as well as the one-loop gauge splitting mentioned above, so that the net splitting remains at 3.55 keV. This gives rise to the following condition
\begin{equation}
2\lambda \langle \eta \rangle+\frac{1}{8\pi^2} \left[g^2_R f(r_R) -g^2_L f(r_L) \right] M_{\Sigma} = 3.55 \; \text{keV}, \;\; \lambda = \frac{\lambda_{\zeta} \langle \zeta \rangle}{\Lambda}
\end{equation}
This fine-tuned cancellation, though seems unnatural, is adopted in this study as it gives rise to interesting dark matter phenomenology.

It should be noted that although we are adopting the scenario (b) described above for generating mass splitting between the two DM candidates, it is non-trivial to forbid the non-zero vev of the neutral components of bitriplet $\psi$. Even if we choose its mass squared term to be positive $\mu^2_{\psi}>0$ to prevent spontaneous generation of non-zero vev, it can still acquire non-zero induced vev after electroweak symmetry breaking. This is due to the existence of trilinear potential terms of the form (ignoring the details of isospin indices)
$$ V^{\psi}_{\text{trilinear}} = \mu_1 \Phi^{\dagger} \psi \Phi + \mu_2 \Delta^{\dagger}_L \psi \Delta_R$$
which can not be forbidden by any symmetry while allowing $\overline{\Sigma_L} \psi \Sigma_R $ at the same time. After electroweak symmetry breaking, the neutral component of $\Delta_L$ acquires a non-zero induced vev as \cite{lrsmpot}
$$v_{L}=\gamma \frac{M^{2}_{W_L}}{v_{R}}$$
where $\gamma$ is a dimensionless parameter given by
\begin{equation}
\gamma = \frac{\beta_2 k^2_1+\beta_1 k_1 k_2 + \beta_3  k^2_2}{(2\rho_1-\rho_3)(k^2_1+k^2_2)}
\label{eq:gammaLR}
\end{equation}
where $\beta, \rho$ are dimensionless parameters of the scalar potential \cite{lrsmpot}. The vev's of bidoublet are denoted by $k_1, k_2$ whereas the vev of the neutral component of $\Delta_R$ is denoted by $v_R$. Since we are considering a vanishing or very suppressed vev of bitriplet, there has to be cancellation between these two terms
$$\mu_1 (k^2_1+ k^2_2) + \mu_2 v_L v_R \approx 0$$
in the absence of any additional symmetries forbidding such trilinear terms. If we consider the bitriplet to be charged under the discrete $Z_N$ symmetry mentioned above, then such terms coupling bitriplet with bidoublet and triplet scalars can be generated through dimension four terms involving another scalar field required to balance the $Z_N$ charges. The same additional field will also appear in generating an effective $\overline{\Sigma_L} \psi \Sigma_R $ term at dimension five level.

Another important issue here is the formation of domain walls if $Z_N$ symmetries are considered which gets broken spontaneously at some stage. These domain walls, if stable on cosmological time scales, will be in conflict with the observed Universe \cite{Kibble:1980mv,Hindmarsh:1994re}. If these walls appear before cosmic inflation, then their density in the present Universe will be too diluted to be of any relevance. But even if we do not assume anything about the scale of inflation, then also one can address the issue of DW by adopting the mechanism suggested by \cite{Rai:1992xw,Lew:1993yt}. The authors in these works considered higher dimensional Planck scale suppressed operators to be the source of domain wall instability which make them disappear. This method was implemented in generic LRSM by the authors of \cite{Mishra:2009mk}. The implementation of this mechanism as a solution to the DW problem in LRSM typically puts an upper bound on the scale of gauge symmetry breaking.

\begin{figure}[t]
\includegraphics[width=0.85\textwidth]{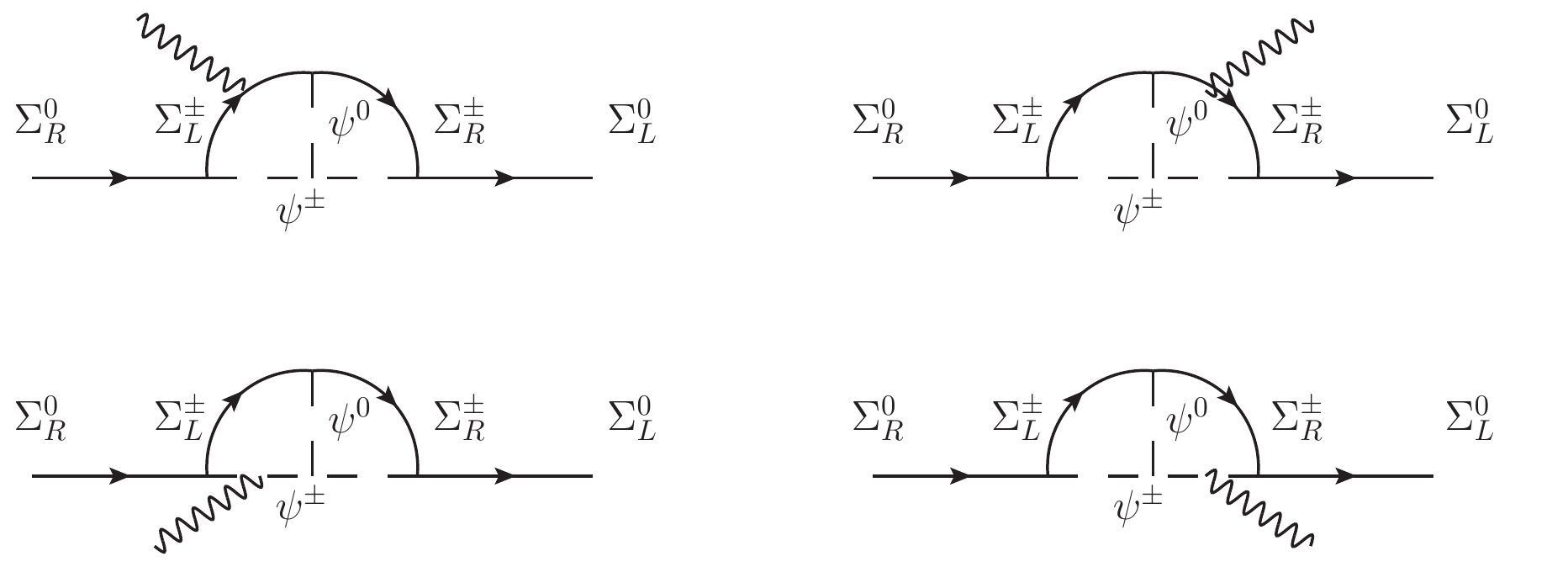}
\caption{Feynman diagram for decay of a heavier fermionic triplet component to 
         light one plus photon as a explanation of $3.55$~keV X-ray line signal.}
\label{sigdecay}
\end{figure}

According the the above analysis, the lightest neutral fermion triplet mass eigenstate is stable and we refer to this state $\Sigma_1$ as stable dark matter candidate. Since $\Sigma_{L}$ does not mix with $\Sigma_R$ in the absence of bitriplet vev, we can denote $\Sigma_1 \equiv \Sigma^0_L, \Sigma_2 \equiv \Sigma^0_R$ and can use them interchangeably. At two loop level the heavier dark matter can decay into the lighter one and a photon through the Feynman diagrams shown in figure \ref{sigdecay}. Here we are adopting the approach $(b)$ to generate the mass splitting and consider the process shown in the figure \ref{sigdecay} is the most dominant one responsible for heavier dark matter decay.
The decay of heavier dark matter component into lighter one plus a monochromatic photon is possible if 
the life-time of the heavier dark matter component is larger than the age of the 
Universe. The same process $\Sigma_2 \rightarrow \Sigma_1 \gamma$ is displayed in 
Fig.\ref{sigdecay} and the decay width expression can be written as
\begin{align}
\Gamma (\Sigma_2 \rightarrow \Sigma_1 \gamma) &\simeq \frac{\left(M^2_{\Sigma_2} - M^2_{\Sigma_1}\right)^3}{16\pi M^3_{\Sigma_{2}}}
\left[F^2_1 + F^2_2 \right]
\end{align}
where,
\begin{align}
F_1 &\simeq 2\left(\frac{M_{\Sigma_{2}}-M_{\Sigma_1}}{m^2_{\psi}}\right)I \; ;& 
F_2 &\simeq 2\left(\frac{M_{\Sigma_{2}}+M_{\Sigma_1}}{m^2_{\psi}}\right)I \; ; &
I &\simeq \left(\frac{Y^3_{\psi}\mu_{\psi}}{256 \pi^4 m_{\psi}}\right)
\label{eq:decayN}
\end{align}
Here, $F_1$ and $F_2$ correspond to Lorentz invariant form factors connected to electric dipole moment transition and purely 
magnetic moment transition respectively. We have assumed that the $CP$ is 
not violated by the interactions involving $\Sigma_{L,R}$ which will lead to two possible $CP$ 
eigenvalues $\in \{+i,-i\}$ for the Majorana particles $\Sigma_R$ and $\Sigma_L$. For same $CP$ eigenvalues we get purely electric dipole moment transition for which $F_2=0$  in eq. \eqref{eq:decayN} and for opposite $CP$ eigenvalues we get purely magnetic moment transition \cite{Kayser:1984ge,Pal:1981rm} for which $F_1=0$. In the above expressions, $\mu_{\psi}$ is the trilinear bitriplet coupling and $m_{\psi}$ is the bitriplet scalar mass. We take them equal to $v_R$ in our calculations.

\begin{figure}[h!]
\includegraphics[width=0.6\textwidth]{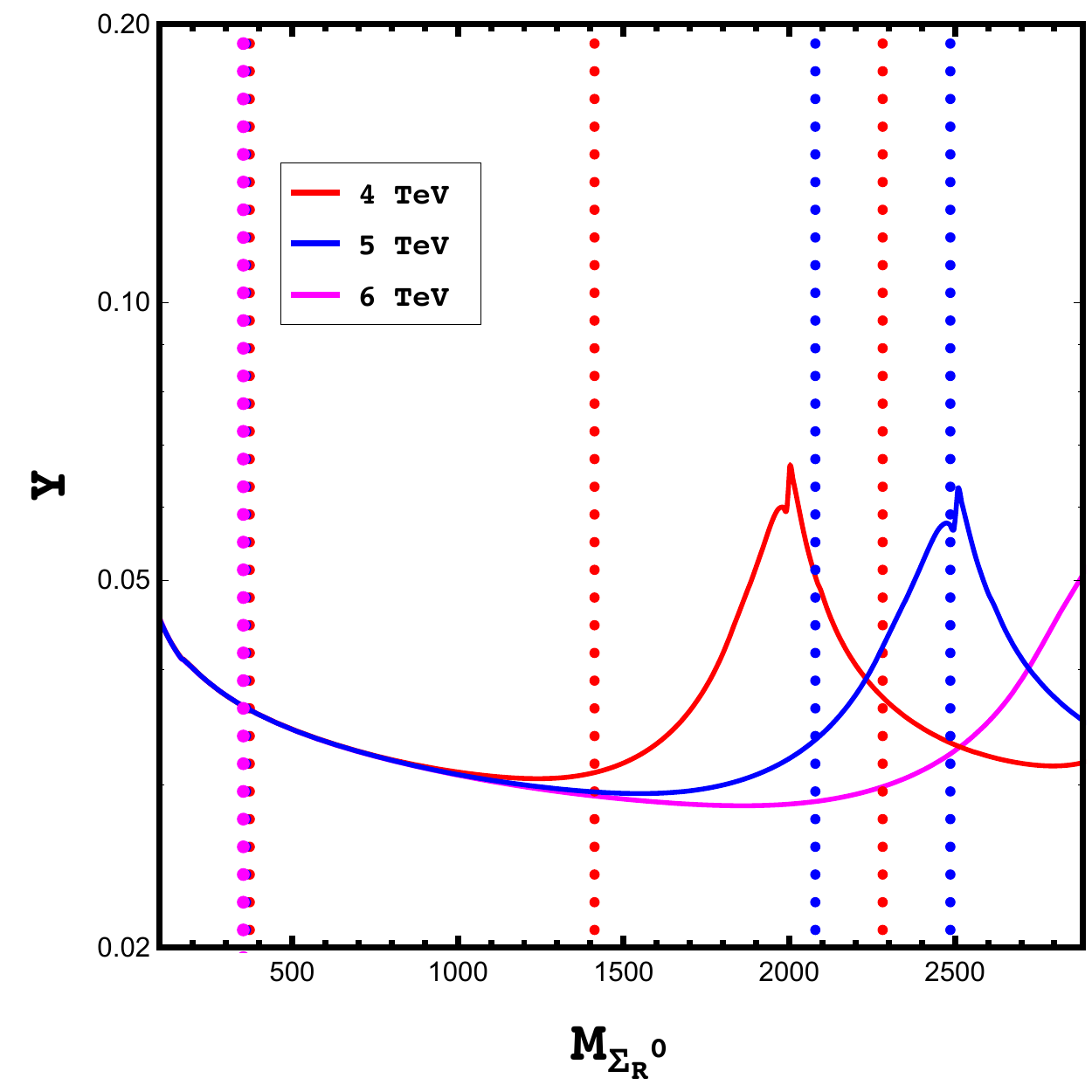}
\caption{Constraint on bitriplet-triplet coupling $Y_\psi$ from X-ray data. The solid lines correspond to the constraints on $Y_\psi-M_{\Sigma^0_R}$ parameter from X-ray data for different values of $W_R$ masses $4, 5, 6$ TeV represented by red, blue and pink respectively.  The vertical dashed lines correspond to the heavier dark matter masses $M_{\Sigma^0_R}$ for which the total dark matter relic abundance is same as the observed one.}
\label{decayplot}
\end{figure}
In order to fit our model with the observed $3.55$ keV X-Ray line data \cite{Xray1}, we follow the constraint on the 
decay width of the heavier dark matter candidate $\Sigma_2$ as obtained in\cite{Falkowski:2014sma}
\begin{equation}
\Gamma_{\Sigma_2 \rightarrow \Sigma_1 \gamma} \approx 6.2 \times 10^{-47} M_{\Sigma_2} \; \text{GeV}
\label{xray}
\end{equation}
where $\Sigma_2$ contributes around $50\%$ to dark matter relic abundance. 
Here the dependence on $M_{\Sigma_2}$ arises via the number density of dark matter. Since the relative contribution to the total dark matter relic abundance need not be $50\%$ in our model, we multiply 
the right hand side of above equation \eqref{xray} by a factor $\frac{1}{2}\frac{\Omega_{\text{DM}}}{\Omega_{\Sigma_2}}$. This is clear from the fact that, lesser the relic abundance of $\Sigma_2$, the more should be the decay width to give rise to the observed flux. Choosing the bitriplet mass and trilinear mass term appearing in the decay width expression to be same as $v_R$ as mentioned above, 
we constrain the dimensionless parameter $Y_\psi=Y$ and dark matter mass from the requirement of satisfying the above constraints on decay width from X-ray data. The constrained parameter space for the CP odd case mentioned above is shown in figure \ref{decayplot}. It can be seen that for TeV scale dark matter masses, one does not require heavy fine-tuning of Yukawa coupling $Y_{\psi}$ to generate the required life-time. The vertical dashed lines in the plot shown in figure \ref{decayplot} corresponds to the masses of heavier dark matter for which the total relic abundance of the two dark matter candidates satisfy the Planck bound \eqref{dm_relic}. The peaks in the plot correspond to the masses at which the relative abundance of the heavier dark matter candidate is minimum due to resonant annihilation, so that one requires a larger value of coupling to enhance the decay width giving rise to the same observed X-ray flux. The relation between these peaks and the dark matter relic abundance will become clear in the next section, when we discuss the calculation of relic abundance.
\begin{figure}[h!]
\centering
\begin{tabular}{c}
\includegraphics[width=0.75\textwidth]{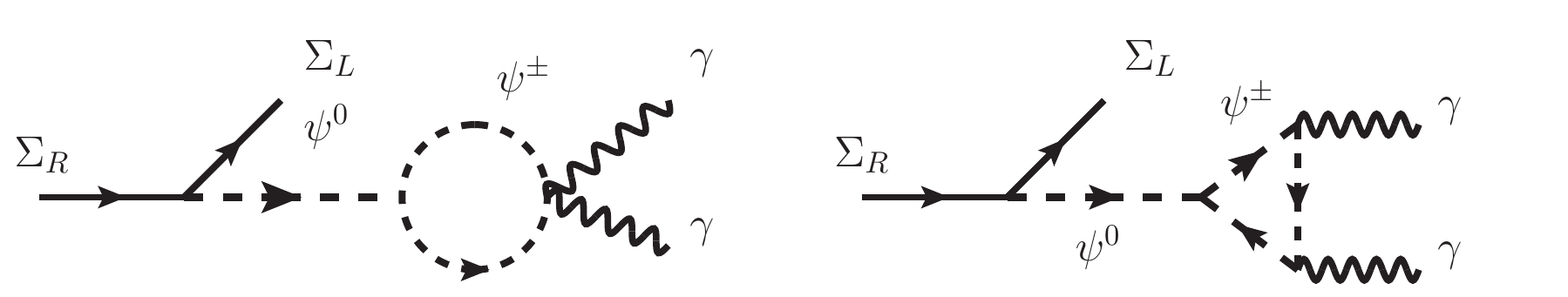}
\end{tabular}
\caption{Three body decay of heavier dark matter into the lighter one and two photons at one loop.}
\label{decayplot1}
\end{figure}

\begin{figure}[h!]
\includegraphics[width=0.6\textwidth]{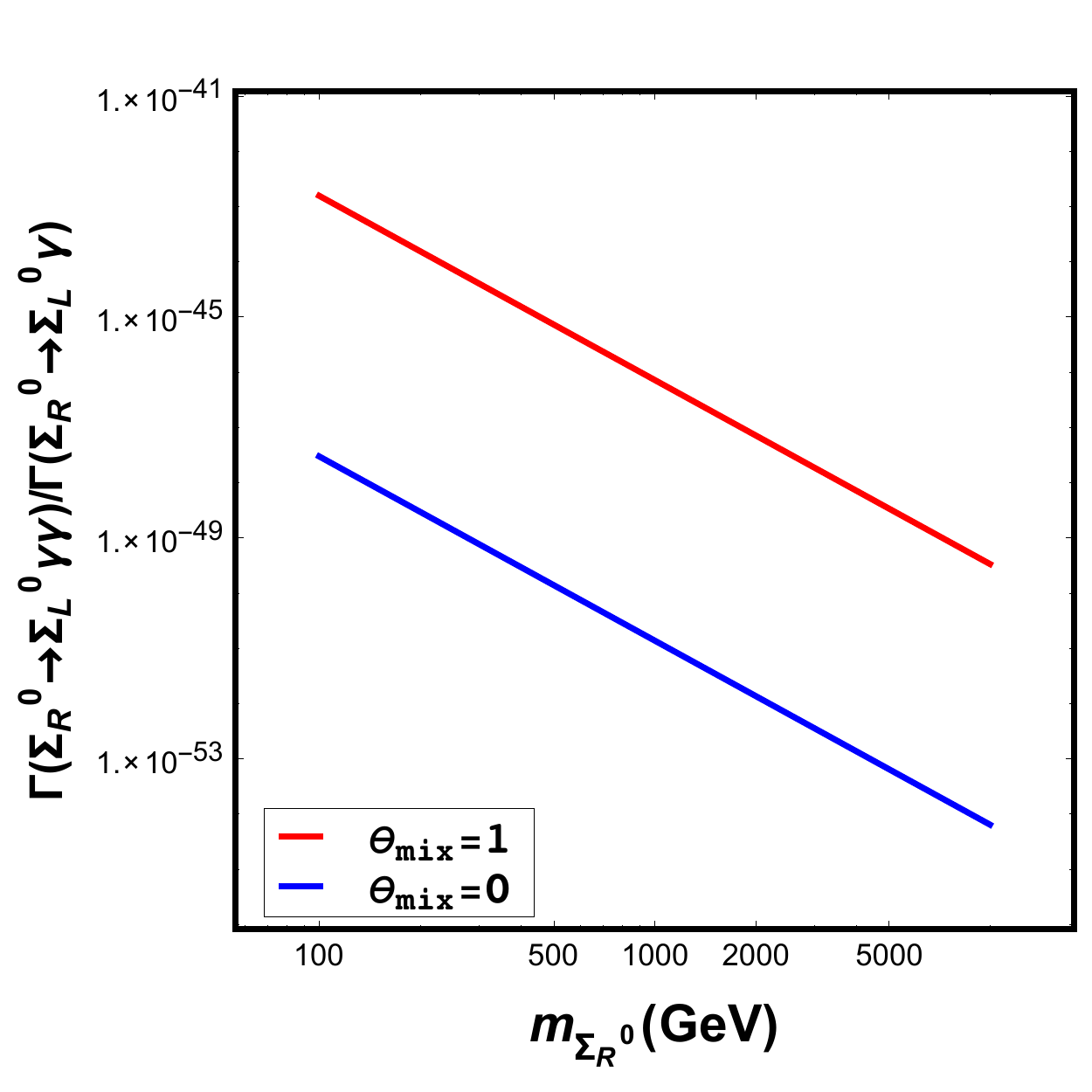}
\caption{Comparison of three body decay at one loop with two body decay at two loop. $\theta_{\text{mix}} =\theta$ is the mixing between bitriplet scalar and SM Higgs discussed in the text.}
\label{decayplot2}
\end{figure}
We also note that there are three body decay diagrams of $\Sigma_2$ into $\Sigma_1$ and two neutrinos or two photons, which may compete with the two loop diagram discussed above. The process $\Sigma_2 \rightarrow \Sigma_1 \gamma \gamma$ can in fact arise at one loop level through an off-shell neutral bitriplet scalar which then can go into two photons at one loop through the charged bitriplet scalars, as shown in figure \ref{decayplot1}. Similarly, one can also have such three body decay due to the mixing between bitriplet scalar and SM Higgs (arising due to trilinear scalar couplings mentioned above) so that the loops giving rise to two photons are mediated by charged particles of the SM. Out of these charged particles, the contribution of the top quark and $W$ boson is significant. Since the amplitude for Higgs decay to diphotons mediated by top quark and $W$ bosons are of the same order (but with opposite sign), here we check the contribution of top quarks only. Showing the three body decay of heavier DM due to top quark loop to be significantly smaller than the two body decay will be enough to make sure that inclusion of $W$ boson loop will also keep the three body decay width smaller. On the other hand, the other process $\Sigma_2 \rightarrow \Sigma_1 \nu_L \nu_L$  can happen at tree level through the neutral bitriplet scalar which can then go into two neutrinos through its mixing with the neutral left handed triplet scalar $\delta^0_L$ through trilinear scalar couplings of the form $\Delta^{\dagger}_L \psi \Delta_R$. Even if the mixing between $\delta^0_L$ and $\psi^0$ is of order unity, the decay width can be suppressed by tiny Yukawa couplings of neutrinos. For example, if $v_L \sim 1$ GeV, then the Yukawa couplings have to be smaller than $10^{-10}$ which will suppress the corresponding decay channel to neutrinos and $\Sigma_1$. This decay width can be estimated as
\begin{align*}
\Gamma_{\Sigma^0_R \rightarrow \Sigma^0_L \nu \nu} 
\simeq \frac{Y^2_\Sigma Y^2_\nu \theta^2 m_{\Sigma^0_L}}{768\pi^3}\frac{\Delta k^6}{m^2_{\Sigma^0_R} m^4_{\psi}}
\end{align*}
where $Y_{\nu}$ is the neutrino Yukawa coupling with $\Delta_L$ and $\Delta k$ is the mass difference between the two DM candidates. The mixing between $\delta^0_L$ and $\psi^0$ is denoted by $\theta$. This remains very much suppressed compared to the required decay width of heavier DM for generating the observed 3.55 keV line. Although this process can be tuned to be small by assuming tiny couplings of $\Delta_L$ to neutrinos, the other three body decay $\Sigma_2 \rightarrow \Sigma_1 \gamma \gamma$ can not be tuned arbitrarily as it involves the same mass and couplings that are involved in the two loop decay diagram discussed above. We therefore, calculate it in details and show it in appendix \ref{appen1}. The three body decay width is coming out to be
\begin{align}
\Gamma_{\Sigma^0_R \rightarrow \Sigma^0_L \gamma \gamma} \simeq \frac{Y^2_{\psi}e^4 }{256\pi^3 m^2_{\Sigma^0_R} } \left[\frac{\Delta k^9 \left(8m^2_t m^2_\psi\theta + m^2_t v\mu \right)^2}{90720\pi^4 m_t^4 v^2 m^8_{\psi}}\right]
\end{align}
where $\mu$ is the trilinear mass term in the scalar potential of $\psi$, $\Delta k$ is the mass difference between the two DM candidates, $\theta$ is the mixing between SM Higgs and bitriplet scalar and $m_t$ is the top quark mass. In the small angle approximation, $\theta$ can be derived as 
$$ \theta \approx \frac{\mu_1 \langle \Phi \rangle}{m^2_{\psi}} $$
For negligible mixing $\theta \approx 0$, this decay width correspond to the width originating from the decay diagrams shown in figure \ref{decayplot1}. We compare this three body decay width with the two body one numerically and show their comparison in figure \ref{decayplot2}. It can be seen from this figure that the three body decay remains suppressed by many order of magnitudes compared to the two body decay process for $\Sigma_2$ masses in the range 100 GeV to 10 TeV and same values of mediator mass $m_{\psi}$ and Yukawa coupling $Y_{\psi}$ required to fit the X-ray data, as shown in figure \ref{decayplot}. This is true even for maximal mixing $\theta \sim 1$ between bitriplet scalar and the SM Higgs. Including $W$ boson loop in decay to two photons in the three body decay will not significantly change the three body decay width and it will still remain many order of magnitudes smaller compared to the dominant two body decay. This makes sure that the two loop and two body decay is the most dominant decay mode of the heavier dark matter candidate in our model, validating the subsequent analysis.

\section{Dark Matter in LRSM}
\label{sec3}
Following the minimal left-right dark matter formalism \cite{Heeck:2015qra,Garcia-Cely:2015quu}, we include two additional 
fermion triplets $\Sigma_L \sim (\vec{3},\vec{1},0,1), \Sigma_R \sim (\vec{1},\vec{3},0,1)$ as mentioned above. 
Due to the absence of any interaction terms mediating the decay of $\Sigma_{L,R}$ into the SM particles, the lightest component 
of each of these triplets are accidentally stable and hence can be CDM candidates if they are electromagnetically neutral. Usually, 
the relic density of CDM is calculated using the formalism discussed in ref. \cite{Jungman:1995df} which can be generalised 
to our present case as already discussed in refs. \cite{Heeck:2015qra,Garcia-Cely:2015quu} as
\begin{equation}
\Omega_{\rm DM} h^2 = \Omega_{\Sigma^0_L} h^2 + \Omega_{\Sigma^0_R} h^2 
\label{eq:relic}
\end{equation}
It should be noted that we are ignoring the coannihilations between the left and right dark matter candidates due to the bitriplet scalar field to be discussed below. Therefore, we are using $\Sigma^0_{L,R}$ to denote the dark matter candidates instead of $\Sigma^0_{1,2}$ in this section. Also, in the absence of coannihilations between left and right sectors, one can independently calculate the relic abundances of left and right sector dark matter candidates.
Thus, the total dark matter relic abundance can be written as $\Omega_{\rm DM} h^2 = \Omega_{\Sigma^0_L} h^2 + \Omega_{\Sigma^0_R} h^2$ which should be in agreement with the 
observed number mentioned in \eqref{dm_relic} given by the Planck experiment~\cite{Planck13}.
\begin{figure}[h!]
\includegraphics[width=0.85\textwidth]{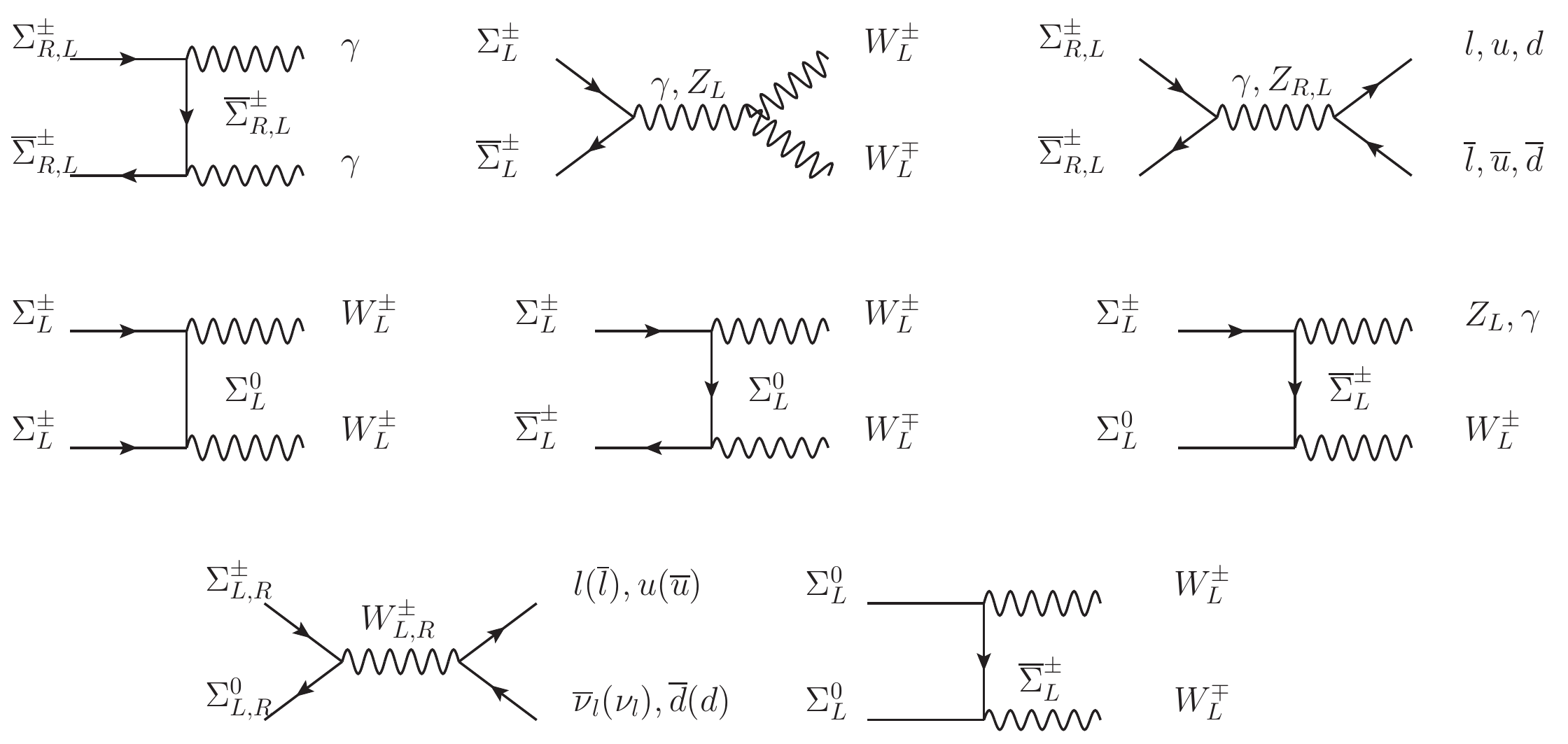}
\caption{The annihilation and coannihilation diagrams for fermion triplet dark matter relevant for relic density computation 
              within the framework of minimal left-right dark matter ~\cite{Heeck:2015qra}. Though there could be additional diagrams 
              with the inclusion of new Yukawa interaction term $\frac{Y_\psi}{2} \overline{\Sigma_L} \psi \Sigma_R$, they have no contributions for vanishing vev of the neutral component of scalar bitriplet 
              $\psi$. All the Feynman diagrams are displayed in terms of $\Sigma_{L,R}$ instead of their mass eigenstates $\Sigma_{1,2}$.}
\label{relicdm}
\end{figure}
However, due to the small mass differences between charged and neutral components of the fermion triplets, there exists co-annihilation channels 
which can affect the dark matter relic abundance.
Before estimating the annihilation and coannihilation channels for triplet fermion dark matter one should know 
the possible interaction terms given by
\begin{align}
\begin{split}
\L_\Sigma &\supset  \big[ g_L    \overline{\Sigma}^+_L \slashed{W}_{L}^3 \Sigma^+_L  + \sqrt{2} g_L  \overline{\Sigma}_L^{+} \slashed{W}^+_L \Sigma^0_L + \hc \big] \\
&+\big[ g_R    \overline{\Sigma}^+_R \slashed{W}_{R}^3 \Sigma^+_R  + \sqrt{2} g_R  \overline{\Sigma}_R^{+} \slashed{W}^+_R \Sigma^0_R+ \hc \big] \, .
\end{split}
\end{align}
Since the high $SU(2)$ dimensions of triplet fermions do not allow them to couple to fermions and scalars, the only interactions 
affecting relic abundance are the gauge interactions from the above kinetic terms. Here the charged component of the fermion triplets 
interact with photon and $Z_{L,R}$ bosons since one can express the weak gauge boson states $W_{L,R}^3, B$ in terms of 
physical gauge boson states. The charged gauge boson mass matrix in $(W^{\pm}_L, W^{\pm}_R)$ basis is given by 
\begin{equation}
M^2_{\pm} =\frac{1}{4} \left(\begin{array}{cc}
g^2_L (k^2_1+k^2_2) & -2g_L g_R k_1 k_2 \\
-2g_L g_R k_1 k_2 & g^2_R (k^2_1+k^2_2) + 2g^2_R v^2_R 
\end{array} \right).
\end{equation}
Here $k_{1,2}/\sqrt{2} = \langle \phi_{1,2}^0 \rangle, v_R/\sqrt{2}=\langle \delta_{R}^0 \rangle$. The physical eigenstates can be found by the orthogonal transformation
\begin{equation}
\left(\begin{array}{c} W^{\pm}_L \\ W^{\pm}_R \end{array}\right)=
\left(\begin{array}{cc}
\cos{\xi} & \sin{\xi} \\
-\sin{\xi} & \cos{\xi}
\end{array} \right)
\left(\begin{array}{c} W^{\pm}_1 \\ W^{\pm}_2 \end{array}\right). 
\end{equation}
where the mixing angle is given by 
\begin{equation}
\tan{2\xi} = -\frac{2g_L k_1 k_2}{g_R v^2_R}
\end{equation}
In the limit $k_2 \rightarrow 0$, this mixing vanishes. Similarly, the neutral gauge boson mass matrix in the $W_{L,R}^3, B$ basis is given by
\begin{equation}
M^2_{0} =\frac{1}{4} \left(\begin{array}{ccc}
g^2_L (k^2_1+k^2_2) & -g_L g_R k_1 k_2 & 0\\
-g_L g_R k_1 k_2 & g^2_R (k^2_1+k^2_2) + 2g^2_R v^2_R & -4g_R g' v^2_R \\
0 & -4g_R g' v^2_R & g'^2 v^2_R 
\end{array} \right)
\end{equation}
which can again be diagonalised by an orthogonal transformation given as
\begin{equation}
\left(\begin{array}{c}W_{L \mu}^3\\W_{R \mu}^3\\B_\mu\end{array}\right)=
\left(\begin{array}{ccc}
c_{W}c_{\phi}    &  c_W s_{\phi}                           &  s_{W} \\
-s_{W} s_M c_{\phi}-c_M s_{\phi}       & -s_W s_M s_{\phi}+c_M c_{\phi}   & c_{W} s_M    \\
-s_W c_M c_{\phi}+s_M s_{\phi}   & -s_W c_M s_{\phi}-s_M c_{\phi}   &  c_W c_M
\end{array} \right)
\left(\begin{array}{c}Z_{L \mu}\\Z_{R \mu}\\A_\mu\end{array}\right). 
\label{neutral1}
\end{equation}
where $c_{W, m, \phi}=\cos\theta_{W, m, \phi}$, $s_{W, m, \phi}=\sin\theta_{W, m, \phi} $. These angles in the limit $k_{1,2} \ll v_R$ are given by 
\begin{equation}
s_{W}  =\frac{g'g_R}{(g^2_L g^2_R + g'^2(g^2_L+g^2_R))^{1/2}}, \;\; s_M = \frac{g_L}{g_R} \tan{\theta_W}, \;\; s_{\phi} \approx 0
\end{equation}
In the limit $g_L =g_R=g$, this corresponds to the usual values $s_W = g'/\sqrt{g^2+2g'^2}, s_M = \tan{\theta_W}$ and the above rotation matrix gets simplified to
\begin{equation}
\left(\begin{array}{c}W_{L \mu}^3\\W_{R \mu}^3\\B_\mu\end{array}\right)=
\left(\begin{array}{ccc}
c_{W}    & 0                           &  s_{W} \\
-s_{W} t_{W}       & \frac{\sqrt{c_{W}^2-s_{W}^2}}{c_{W}}   & s_{W}    \\
-t_{W} \sqrt{c_{W}^2-s_{W}^2}    & -t_{W}                                  & \sqrt{c_{W}^2-s_{W}^2} 
\end{array} \right)
\left(\begin{array}{c}Z_{L \mu}\\Z_{R \mu}\\A_\mu\end{array}\right)
\end{equation}
where $t_{W}=\tan\theta_W$.

Although we are denoting the two dark matter states as $\Sigma_{1,2}$ to adopt generality, in the absence of mixing between $\Sigma^0_L, \Sigma^0_R$, they are identical except for a tiny mass splitting of 3.55 keV. We know that the electromagnetic coupling 
between $\Sigma_1$, $\Sigma_2$ and $\gamma$ has to be very much suppressed from the requirement of decaying dark matter 
to give $3.55~$keV X-ray line signal $\tau(\Sigma_2 \to \Sigma_1 \gamma)$. 
The suppressed coupling makes it difficult to bring $\Sigma_1$ and $\Sigma_2$ into thermal equilibrium with each other and thus, the calculation 
of relic abundance will be similar to the formalism discussed in refs~\cite{Heeck:2015qra,Garcia-Cely:2015quu}, as discussed above. Additionally, the interactions with scalar 
bitriplet can also affect the coannihilations between left and right components of dark matter. Among bitriplet-triplet interactions, the most 
important co-annihilation channel is $\Sigma^0_{1} \Sigma^0_{2} \rightarrow W^+_L W^-_L $ through s-channel exchange of neutral triplet $\epsilon^0$. However, if the bitriplet does not acquire any vev (in the approach $(b)$ above that we adopt here), this particular annihilation channel is absent since the $W^+_L W^-_L \epsilon^0$ vertex is proportional 
to $g^2_2 \langle \epsilon^0 \rangle $. Therefore, the relic abundance of both left and right handed triplet dark matter candidates can be calculated independently. 
However, there exists coannihilation channels between charged and neutral component of a fermion triplet within individual left and right sectors which are taken into account. In principle, there can also be a coannihilation channel where the two dark matter candidates can go into neutral bitriplet scalars. However, this is kinematically forbidden due to the choice of bitriplet mass equal to $v_R$ (which is larger than dark matter mass satisfying total relic abundance) as discussed in the previous section. The (co)annihilation channels included in the calculation of dark matter relic abundance are shown in figure \ref{relicdm}.

\begin{figure}[t!]
\begin{tabular}{c}
\includegraphics[width=0.5\textwidth]{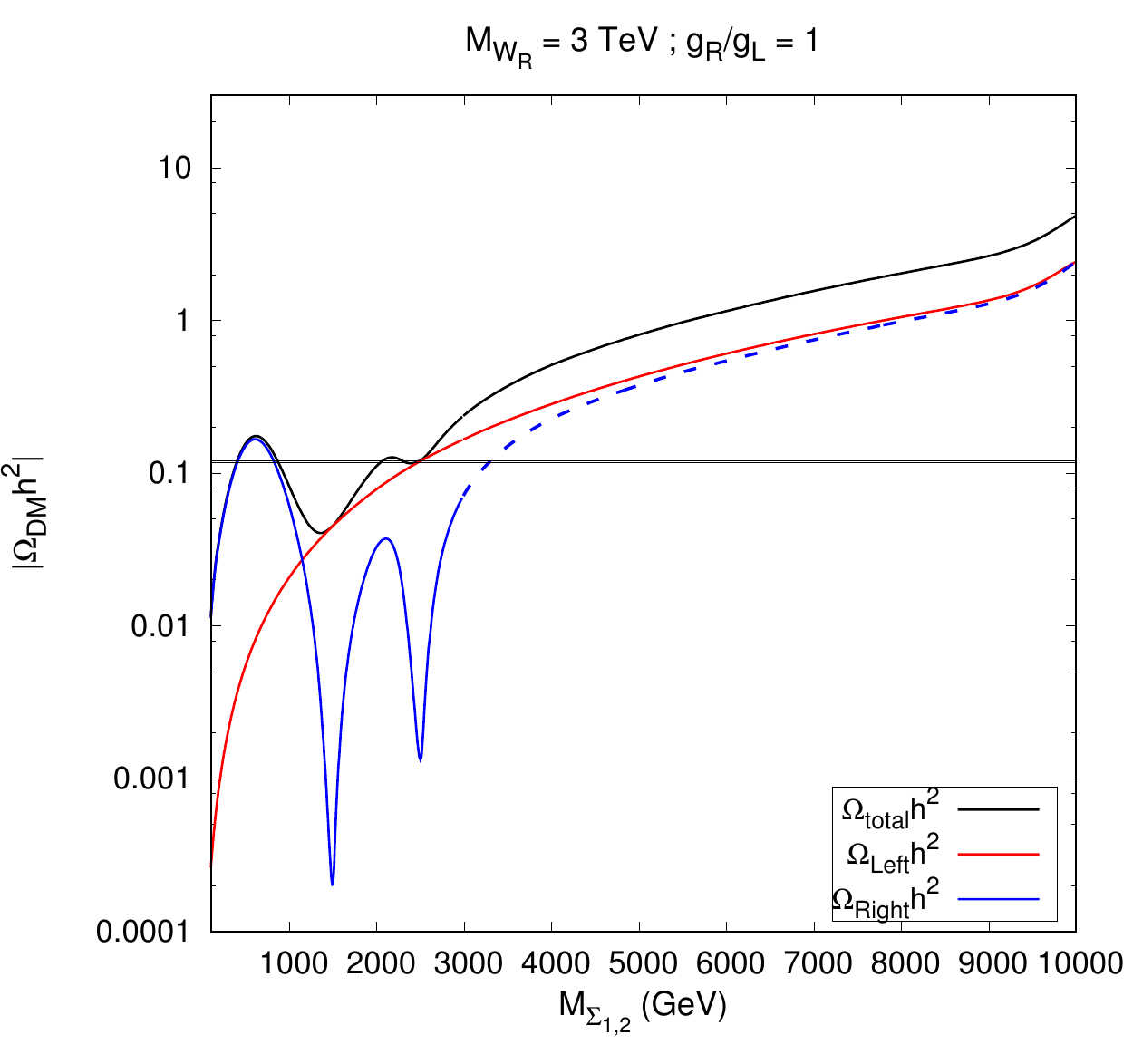}
\includegraphics[width=0.5\textwidth]{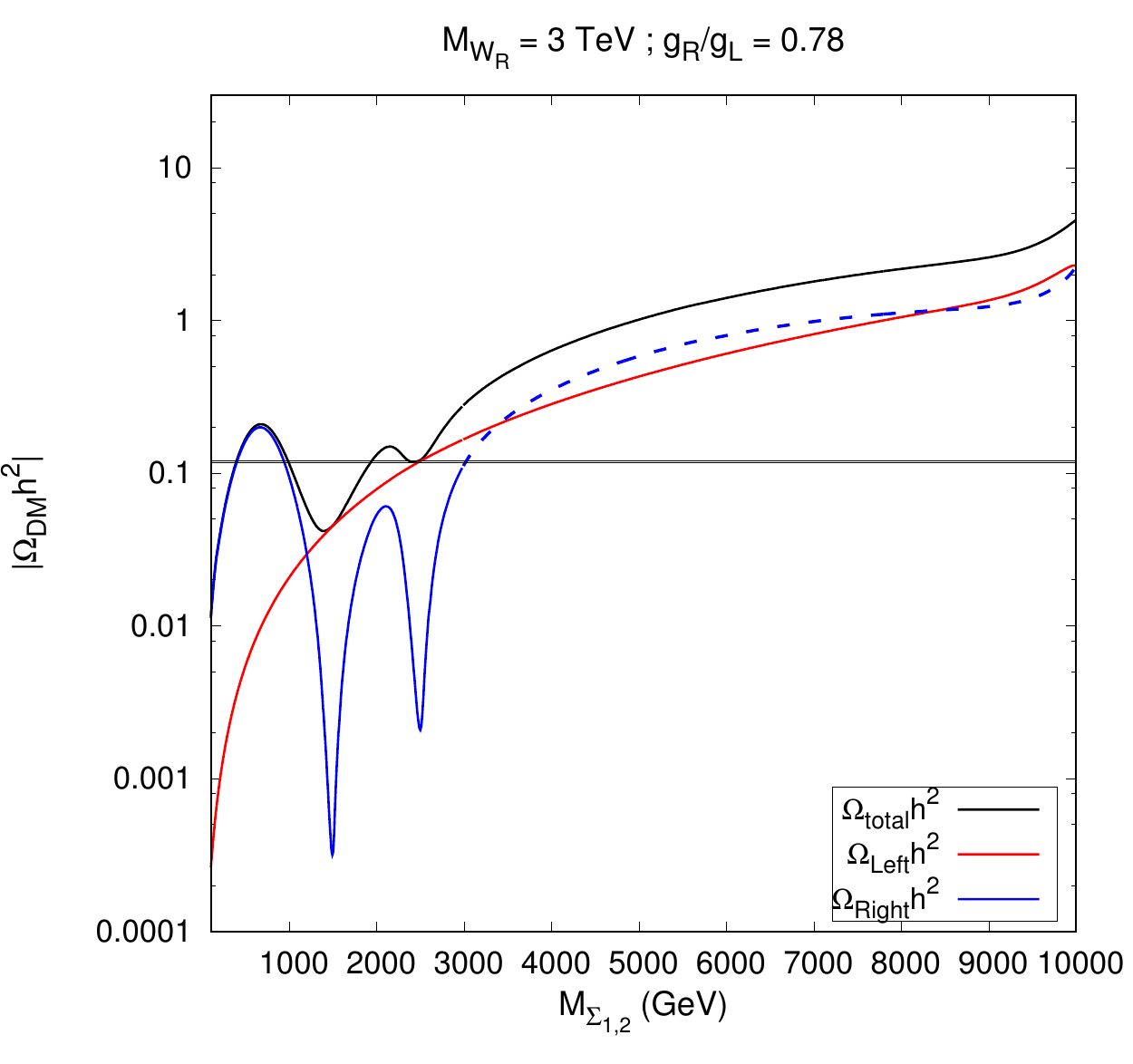}
\end{tabular}
\caption{Relic abundance of dark matter as a function of dark matter mass for $M_{W_R}=3$ TeV. The horizontal black line corresponds to the Planck limit on dark matter abundance. The dashed part of the blue curve corresponds to the region where charged component of right fermion triplet becomes lighter than the neutral one.}
\label{relicdm1}
\end{figure}
This effects of coannihilation on dark matter relic abundance were studied by several authors in \cite{Griest:1990kh, coann_others}. Here we follow 
the analysis of \cite{Griest:1990kh} to calculate the effective annihilation cross section in such a case. As the coannihilation processes through mediation 
of bitriplet scalar is negligible, the total abundance is effectively a sum of the contributions coming from left and right fermion triplets. 
Using the annihilations and coannihilation 
processes, the formula for relic abundance for dark matter candidates $\Sigma^0_L$ and $\Sigma^0_R$ 
is given by
\begin{align}
\Omega_{\rm DM} h^2 &=\Omega_{\Sigma_{1}} h^2 + \Omega_{\Sigma_{2}} h^2 \simeq \Omega_{\Sigma_{L}} h^2+ \Omega_{\Sigma_{R}} h^2\,, \nonumber \\
\Omega_{\Sigma_{A}} h^2 &= \frac{1.09 \times 10^{9}\, \mbox{GeV}^{-1}}{\sqrt{g^*} M_{\rm Pl}} \frac{1}{J(x_F)}
\end{align}
where $A=L,R$ and the factor $J(x_F)$ is defined as follows
\begin{align}
J(x_F) =\int^{\infty}_{x_F} \frac{\langle \sigma|v \rangle_{\rm \Sigma_{A}}}{x^2}
\end{align}
In the present work with fermion triplet dark matter, the effective annihilation cross-section for the fermion triplet can be written as 
\begin{align}
&\langle \sigma|v \rangle_{\Sigma_{A}} = \frac{g^2_{\Sigma^0_A}}{g^2_{\text{eff}}} \sigma(\Sigma^0_A \Sigma^0_A) 
        +4 \frac{g_{\Sigma^0_A}  g_{\Sigma^\pm_A} }{g^2_{\text{eff}}} \sigma(\Sigma^0_A \Sigma^\pm_A) \left(1+\Delta_A \right)^{3/2} 
        \mbox{Exp}(-x \Delta_A)+\frac{g^2_{\Sigma^\pm_A}}{g^2_{\text{eff}}} \big(2\sigma(\Sigma^\pm_A \Sigma^\pm_A)  \nonumber \\
& +2 \sigma(\Sigma^+_A \Sigma^-_A) \big )\left(1+\Delta_A \right)^{2} \mbox{Exp}(-2x \Delta_A)
\end{align}
where $\Delta_A=(M_{\Sigma^\pm_A}-M_{\Sigma^0_A})/M_{\Sigma^0_A}$ is the mass splitting ratio and $x=M_{\Sigma^0_A}/T$. Here $A=L,R$ denotes the dark matter candidates. Using ref.~\cite{Heeck:2015qra,Garcia-Cely:2015quu}, the mass splittings between charged and neutral components for left-handed 
and right-handed triplet fermions are given by
\begin{align}
\begin{split}
M_{\Sigma_L^\pm}-M_{\Sigma_L^0} &\simeq  \alpha_2  M_{W} \sin^2 (\theta_W/2) + \mathcal{O}(M_{W}^3/M^2_\Sigma) \,,
\end{split}
\end{align}
\begin{align}
M_{\Sigma_R^\pm}-M_{\Sigma_R^0} &\simeq \frac{\alpha_2}{4\pi} \frac{g_R^2}{g_L^2} M \left[ f(r_{W_2}) - c_M^2 f(r_{Z_2})-s_W^2 s_M^2 f(r_{Z_1})- c_W^2 s_M^2 f(r_\gamma)\right] .
\label{eq:RH_mass_splitting}
\end{align}
Here the one loop self-energy corrections through mediations of gauge bosons are presented within the square bracket of the second expression. For example,  the mass splitting with the approximation $M_\Sigma \gg M_{W_R}$ goes as $\alpha_{2} \left( M_{W_R} - c_M^2 M_{Z_R}\right)/2$. The sine and cosine of different angles $c_M, c_W, s_M, s_W$ etc. correspond to the angles involved in the rotation of neutral gauge bosons given in \eqref{neutral1}. Also, the loop function $f(r)$ is same as the one given in \eqref{eq:fr}.

\begin{figure}[t!]
\begin{tabular}{c}
\includegraphics[width=0.5\textwidth]{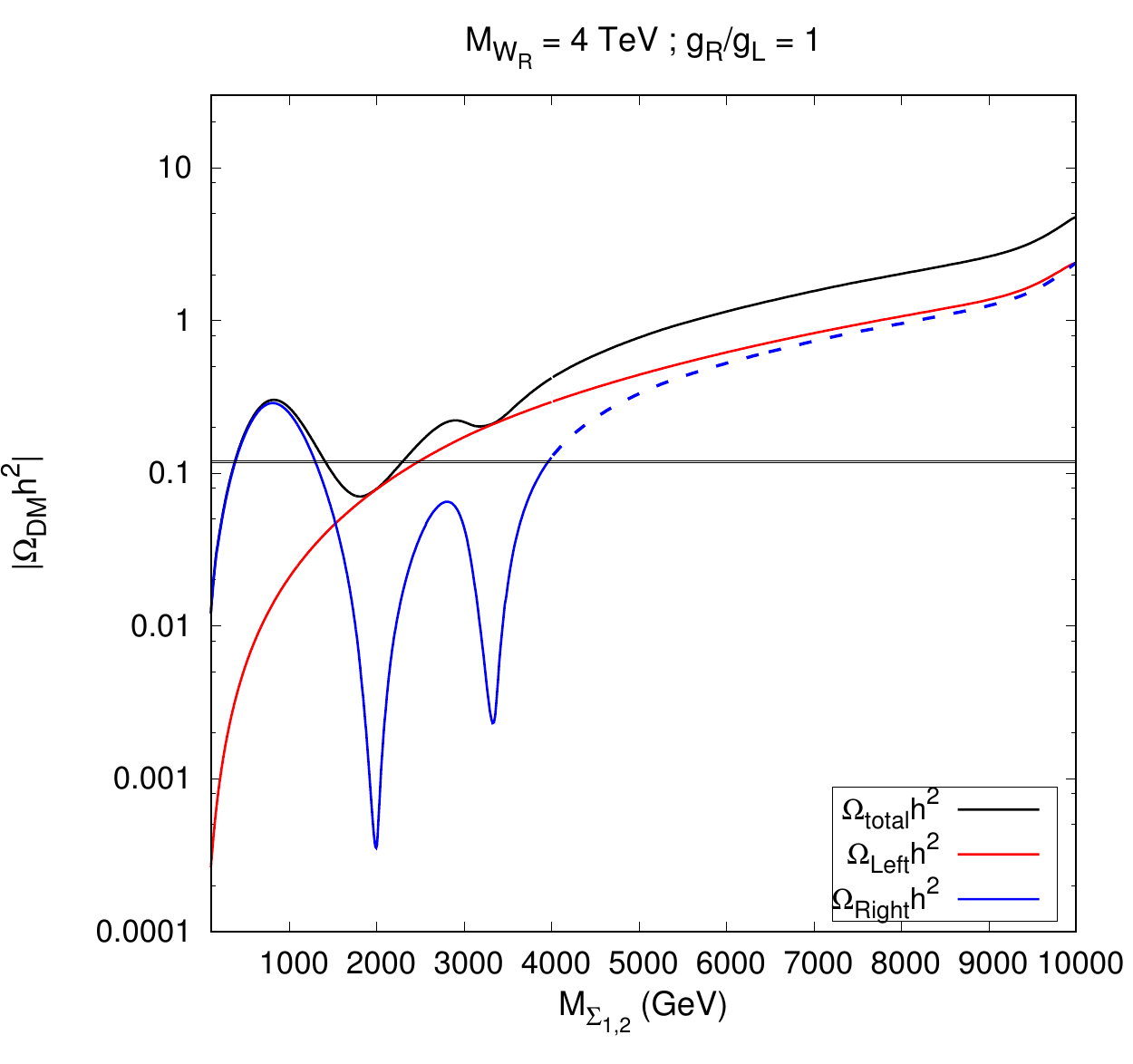}
\includegraphics[width=0.5\textwidth]{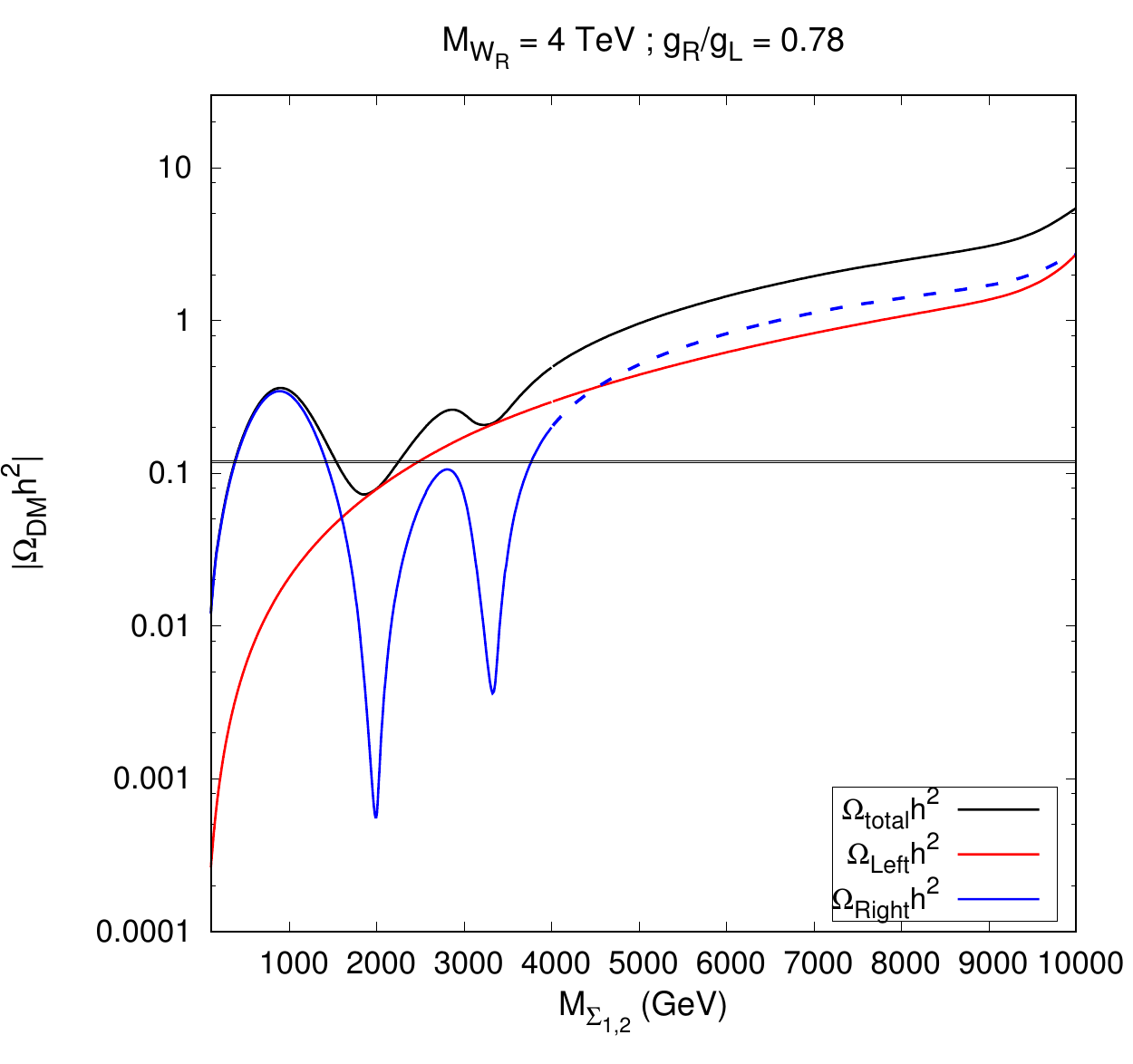}
\end{tabular}
\caption{Relic abundance of dark matter as a function of dark matter mass for $M_{W_R}=4$ TeV. The horizontal black line corresponds to the Planck limit on dark matter abundance. The dashed part of the blue curve corresponds to the region where charged component of right fermion triplet becomes lighter than the neutral one.}
\label{relicdm2}
\end{figure}
Also we denote other parameters like 
$g_{\text{eff}}$ as effective relativistic degrees of freedom while $g_{\Sigma_A} $ and  $g_{\Sigma^\pm_A} $ with $A=L,R$ are 2 for the neutral components of fermion triplets. 
The effective relativistic degrees of freedom is related to other spin degrees of freedom as 
\begin{align}
g_{\text{eff}}=g_{\Sigma^0_A} +2 g_{\Sigma^\pm_A} \left(1+\Delta_A \right)^{3/2}  \mbox{Exp}(-x \Delta_A) 
\end{align}
However, as stressed earlier we do not consider any coannihilation 
between left and right fermion dark matter which allows us to compute the abundance of left and right sector dark matter independently. While calculating 
the relic abundance, we keep the mass difference between left and right handed fermion triplet to be 3.55 keV as required to explain the monochromatic 
X-ray line. The resulting relic abundance for $M_{W_R}=3, 4$ TeV are shown in figure \ref{relicdm1} and \ref{relicdm2} respectively. It is seen from both 
the figures that left-handed triplet dark matter satisfies Planck bound \eqref{dm_relic} for mass around 2.5 TeV, which is very close to the results obtained earlier by the authors of \cite{Ema2009}. The right handed dark matter abundance 
remains suppressed beyond 1.5 TeV. The dips in the right handed dark matter line around $M_{\Sigma_{2}} \approx M_{W_R}/2$ and $M_{\Sigma_{2}} \approx M_{Z_R}/2$ come from the $W_R$ and $Z_R$ resonances respectively. These gauge bosons mediate the $\Sigma^{\pm}_2, \Sigma_2$ and $\Sigma^{+}_2, \Sigma^{-}_2$ annihilations respectively. 
In the case of $M_{W_R} = 3$ TeV, we do not get right handed dark matter abundance beyond  $M_{\Sigma_{2}} \approx  3$ TeV as the charged component 
of right handed triplet becomes lighter than the neutral one for that region of parameter space. For the same reason, we do not get correct relic abundance for right fermion triplet dark matter mass beyond 4 TeV if $M_{W_R} = 4$ TeV. This is highlighted as the dashed lines in the plots shown in figure \ref{relicdm1} and \ref{relicdm2}. The total relic abundance is shown as the black solid line in both these plots. It is clear that for these two benchmark values of $W_R$ masses, the total relic abundance can satisfy the observed dark matter abundance even for dark matter mass as small as a few hundred GeVs. We also show the difference in results for different choices of $g_R$, the $SU(2)_R$ gauge coupling constant. The left panel plots of figure \ref{relicdm1} and \ref{relicdm2} show the results for $g_R=g_L$, in the left-right symmetric limit. The right panels of the corresponding figures show the results for a value of $g_R$ less than $g_L$. This slightly shifts the relic line for right fermion triplet DM upwards due to the decrease in its gauge annihilation channels, as expected. For illustrative purposes we choose $g_R = 0.78 g_L$, as suggested from gauge coupling unification at high energy scale, to be discussed below. Our observations agree with the results obtained by the authors of \cite{Heeck:2015qra,Garcia-Cely:2015quu}.

In the simple relic density computation of two component fermionic dark matter presented in this work, we have not taken into account the Sommerfeld enhancement. However, the detailed discussion of Sommerfeld effects on the abundance of dark matter can be found in \cite{Garcia-Cely:2015quu}. Such non-perturbative effects arise due to exchange of gauge bosons among non-relativistic dark matter particles that are much heavier than the gauge boson masses. This leads to an enhancement in the dark matter annihilation cross sections thereby reducing the dark matter abundance, compared to the one obtained from perturbative calculations. Such Sommerfeld effects on dark matter abundance were also discussed in earlier works~\cite{Hisano:2003ec,Hisano:2004ds, Hisano:2006nn}. Since in this work, we are sticking to right fermion triplet mass comparable or smaller than the respective gauge boson masses, these effects are suppressed for right fermion triplet dark matter. However, for left triplet fermion dark matter such effects can be significant if the dark matter mass is a few TeV. These effects were calculated by the authors of \cite{Hisano:2006nn} and were also taken into account in \cite{Heeck:2015qra,Garcia-Cely:2015quu} for the left fermion triplet. As seen from the results of \cite{Hisano:2006nn} the difference between the relic abundance with and without incorporating Sommerfeld's enhancement is minimal for dark matter mass below 1 TeV. Therefore, for dark matter mass around 1 TeV where the dark matter sector is mostly composed of right fermion triplet, our estimates for left fermion triplet abundance is approximately valid. Taking the Sommerfeld's enhancement into account will further reduce the abundance of left triplet dark matter in this mass regime, as shown in \cite{Hisano:2006nn}. However, a detailed calculation of such non-perturbative effects is beyond the scope of this present work and hence we note here that our relic abundance calculation is more accurate in the low mass regime. This also helps us to avoid the constraints coming from indirect detection limits on gamma ray searches. For example, the left fermion triplet annihilations into $W^+_L W^-_L, \gamma \gamma$ final states can be enhanced by the same Sommerfeld's enhancement. However, as long as we stick to the dark matter mass around a TeV, these effects are suppressed and left fermion triplet remains under-abundant which help us to get rid of such bounds. This fact was also observed by the authors of \cite{Garcia-Cely:2015quu} where they found that a low mass right fermion triplet dark matter mass below 1 TeV is still consistent with indirect detection constraints. Future indirect detection experiments searching for gamma rays may be able to probe this low mass region whereas the prospects of direct detection remains very weak due to the absence of tree level elastic scattering cross section of right fermion triplet dark matter with nucleons.

\section{Gauge Coupling Unification}
\label{sec4}
In earlier discussions we have shown that adding a pair of fermion triplets and a scalar bitriplet 
to the minimal left-right symmetric theory can explain the $3.55$~keV X-ray line signal 
seen by XMM-Newton X-ray observatory via the decay of next-to-lightest neutral fermion triplet 
to the lightest one plus a photon. In this section, our main focus is to examine whether 
we can successfully embed the framework within a non-SUSY $SO(10)$ GUT while achieving gauge 
coupling unification for three fundamental forces. 
It is already pointed out that unification of gauge couplings is possible 
in left-right symmetric model where the left-right discrete symmetry is either broken at very 
high scale \cite{Chang:1983fu,Borah:2013lva} or at TeV scale leading to LHC scale phenomenology 
\cite{Deppisch:2014zta,Patra:2014goa,Awasthi:2013ff}. Also the gauge coupling unification 
in left-right symmetric model with TeV scale stable left-right fermion triplet dark matter 
was studied in a recent work \cite{Dev:2015pga}. Also stable candidate of dark matter within the context of $SO(10)$ GUT has been discussed in Refs.~\cite{Kadastik:2009cu,Kadastik:2009dj,Frigerio:2009wf, Mambrini:2015vna,Nagata:2015dma,Arbelaez:2015ila,Boucenna:2015sdg}. In the present left-right symmetric model, we have both left-handed as well as 
right-handed fermion triplet dark matter with a tiny mass splitting.
Here we intend to examine gauge coupling 
unification by embedding the present framework in a non-SUSY SO(10) GUT where we have introduced 
the fermion triplet pair as well as a scalar bitriplet to the minimal LRSM for explanation 
of $3.55$~keV X-ray line. 

The embedding of the framework within a non-SUSY $SO(10)$ GUT comes 
with the symmetry breaking pattern as follows
\begin{align}
SO(10) \mathop{\longrightarrow}^{\langle (1,1,15) \rangle \subset 45_H}_{M_U} \mathcal{G}_{2213}  
       \mathop{\longrightarrow}^{\langle \Delta_R \rangle \subset 126_H}_{M_R} \mathcal{G}_{213} 
       \mathop{\longrightarrow}^{\langle \phi \rangle \subset 10_H}_{M_Z} \mathcal{G}_{13}. 
\end{align}
%

 
At first stage of symmetry breaking, $SO(10)$ breaks down to LRSM $\mathcal{G}_{2213}\equiv SU(2)_L \times 
SU(2)_R \times U(1)_{B-L} \times SU(3)_C$ at $M_U$ by assigning a non-zero VEV to Pati-Salam multiplet 
$\langle (1,1,15) \rangle \neq 0 \subset 45_H$. The subsequent stage of symmetry breaking $\mathcal{G}_{2213} 
\rightarrow \mathcal{G}_{SM}$ is done with giving a non-zero VEV to $\Delta_R(1,3,2,1) \subset 126_H$ at $M_R$. The 
LRSM symmetry breaking scale i.e, $M_R$ has been fixed around $10$ TeV so that the resulting $W_R$ mass satisfies 
current LHC limit. The final stage of symmetry breaking is happened with SM Higgs doublet contained in $10_H$. 
In the present framework, we have already fermion triplets $\Sigma_L$, $\Sigma_R$ and a scalar bitriplet $\psi$ 
at $M_R$. An additional colored scalar field $\xi(1_L,3_R,-2/3_{BL},\overline{6}_C)$ is added to assist in exact unification of gauge couplings and 
the particle content of the full model is presented in Table\,\ref{tab4}.

\begin{table}[h!]
\begin{center}
\begin{tabular}{|c|c|c|}
\hline
Group $G_{I}$  & Fermions      & Scalars       \\
\hline \hline
$\begin{array}{l}
G_{213} \\
(M_Z \leftrightarrow M_{R})  \end{array}$  & 
$\begin{array}{l}
Q_{L}(2,1/6,3) \\
u_{R}(1,2/3,3), d_{R}(1,-1/3,3) \\
\ell_{L}(2,-1/2,1), e_{R}(1,-1,1)
  \end{array}$ &
$\begin{array}{l}
\phi(2, \frac{1}{2},1) \end{array}$
 \\
\hline 
$\begin{array}{l}
{\small G_{2213}}\\
(M_{R} \leftrightarrow M_{U})  \end{array}$   &
${\small \begin{array}{l}
q_L(2,1,1/3,3),q_R(1,2,1/3,3) \\
\ell_L(2,1,-1,1),\ell_R(1,2,-1,1) \\ 
\Sigma_L(3,1,0,1), \Sigma_R(1,3,0,1)  \end{array}}$ &
${\small \begin{array}{l}
\Phi (2,2,0,1), \psi(3,3,0,1)   \\
\Delta_R(1,3,2,1), \xi(1,3,-2/3,6)   \end{array}}$                                                                   
\\
\hline
\hline
\end{tabular}
\caption{The fermions and scalars are presented at different stages of symmetry breaking scales i.e, in the mass range $M_Z-M_R$ and $M_R-M_{U}$ 
              where $M_Z$ is the SM $Z$ boson mass, $M_R$ is the scale at which left-right symmetry breaks and $M_U$ is the unification scales where all 
              three gauge couplings unify. We found that gauge coupling unification is possible by introducing an additional scalar $\xi(1_L,3_R,-2/3_{BL},\overline{6}_C)$ 
              (as mentioned in the text) apart from the minimal particle content required for dark matter phenomenology.}
\label{tab4}
\end{center}
\end{table}
The Higgs multiplets responsible for masses of all Standard Model fermions is derived from the decomposition, $16 \otimes 16= 10_s + 120_a + 126_s$. 
The SM Higgs doublet contained in $10_H$ which is decomposed as 
\begin{eqnarray}
\hspace*{-0.5cm}10_H(\Phi) &=& \Phi (2,2,1) \oplus (1,1,6) \quad \quad \mbox{Under Pati-Salam Decomposition}\, \nonumber \\
     &=& \Phi (2,2, 0,1) \oplus (1,1,-1/3, 3) \oplus (1,1,1/3,\overline{3}) \quad \quad \mbox{Under $\mathcal{G}_{2213}$ Decomposition.} \nonumber
\end{eqnarray}
The left-right symmetry is spontaneously broken down to SM by assigning a non-zero VEV to right-handed scalar triplet $\Delta_R$ contained 
in $126_H$ of $SO(10)$ as follows
\begin{eqnarray}
\hspace*{-0.3cm}126_H(\Delta) &=& \Delta_L(3,1,10)+\Delta_R(1,3, \overline{10}) + (2,2,15) + (1,1,6)  \quad \mbox{Under Pati-Salam Decomposition}\, \nonumber \\
     &=& \Delta_L (3,1,2,1) \oplus \Delta_R(1,3,2,1) + \cdots  \quad \quad \mbox{Under $\mathcal{G}_{2213}$ Decomposition.}  \nonumber 
\end{eqnarray}
The $SO(10)$ breaks down to left-right symmetric model along with breaking D-parity spontaneously by assigning a non-zero VEV to $\langle (1,1,15) \rangle \neq 0 \subset 45_H$ 
under the following $SO(10)$ decomposition,
\begin{eqnarray}
\hspace*{-0.3cm}45_H(\zeta) &=& \zeta(1,1,15)+(3, 1, 1)+(1, 3, 1)+(2, 2, 6)  \quad \mbox{Under Pati-Salam Decomposition.}\,  \nonumber 
\end{eqnarray}
The usual quarks and leptons constained in $16$ dimensional spinorial representation of $SO(10)$ as 
\begin{eqnarray}
\hspace*{-0.3cm}16_F &=& F_L(2,1,4)+(1, 2, \overline{4}) \quad \mbox{Under Pati-Salam Decomposition}\, \, \nonumber \\
     &=& q_L (2,1,1/3,3) \oplus q_R(1,2,1/3,3) \nonumber \\
     & & \oplus \ell_L (2,1,-1,1) \oplus \ell_R(1,2,-1,1)   \quad \quad \mbox{Under $\mathcal{G}_{2213}$ Decomposition.} 
\end{eqnarray}
The fermion triplets with $B-L=0$ are contained in $45_F$ of $SO(10)$ with the following decomposition,
\begin{eqnarray}
\hspace*{-0.3cm}45_F &=& (1,3,1)+(1,1,3)+(6,2,2)+(15,1,1) \quad \mbox{Under Pati-Salam Decomposition}\, \, \nonumber \\
     &=& \Sigma_L (3,1,0,1) \oplus \Sigma_R(1,3,0,1) + \cdots   \quad \quad \mbox{Under $\mathcal{G}_{2213}$ Decomposition.} 
\end{eqnarray}
Similarly, the bitriplet scalar is present in $54_H$ representation of $SO(10)$ and the extra scalar assisting gauge coupling unification is contained in $126_H$. 
The scalar singlet $\eta$ which is odd under D-parity is contained in $210_H$ of $SO(10)$ representation. 

In order to get mass of lighter particles around TeV scale while keeping other unwanted particles at GUT scale, one has to follow the mechanism discussed 
in Refs.~\cite{Chang:1983fu, Chang:1984uy, Chang:1984qr}. In this idea of spontaneous D-parity breaking mechanism, an asymmetry between left-handed and 
right-handed scalar fields is introduced leading to coupling constants of $SU(2)_R$ and $SU(2)_L$ evolve separately under the renormalization group resulting 
unequal $SU(2)_L$ and $SU(2)_R$ gauge couplings. As a result of this, the desired particles get their masses at much lower scale i.e, at around TeV range while 
other unwanted particles get their masses around D-parity breaking scale. Since D-parity spontaneously breaks at GUT scale, the unwanted particle have their 
masses close to GUT scale.

The explicit decomposition of $SO(10)$ Higgs representation under left-right symmetric group and the corresponding Clebsch-Gordan coefficients can be found 
in ref.~\cite{Fukuyama:2004ps}. To illustrate the idea of spontaneous D-parity breaking, consider a scalar singlet $\eta$ which is odd under discrete D-parity and two fermion 
triplets $\Sigma_L, \Sigma_R$. The renormalizable term connecting $\Sigma-\eta$ is $\lambda_\Sigma \sigma \mbox{Tr}\left(\Sigma^c_L \Sigma_L - \Sigma^c_R \Sigma_R\right)$ where 
$\lambda_\Sigma$ is the dimensionless coupling and $\mu_\Sigma$ is the bare mass term for fermion triplets i.e, $\mu_{\Sigma} \mbox{Tr}\left(\Sigma^c_L \Sigma_L+ 
\Sigma^c_R \Sigma_R\right)$. Once the D-parity odd singlet scalar takes non-zero VEV, the left-right symmetry with D-parity $SU(3)_c \times SU(2)_L \times SU(2)_R \times U(1)_{B-L} \times D$ 
is spontaneously broken but the gauge symmetry $SU(3)_c \times SU(2)_L \times SU(2)_R \times U(1)_{B-L}$ remains unbroken resulting in 
\begin{subequations}
\begin{eqnarray}
&&M_{\Sigma_R}=\mu_{\Sigma} - \lambda_{\Sigma} \langle \sigma \rangle \, ,  \\
&&M_{\Sigma_L}=\mu_{\Sigma} + \lambda_{\Sigma} \langle \sigma \rangle \, ,
\end{eqnarray}
\end{subequations}

Assigning the large parity breaking vev to the D-parity odd scalar singlet around GUT scale $\langle \sigma \rangle \sim M_{U}$ it can be demonstrated that the left-handed scalars to have heavy masses 
i.e., $\mathcal{O}(M_{U})$ whereas the right-handed scalars can have much lighter masses near the TeV scale with $M_{\Sigma_R} \simeq \left(\mu_{\Sigma} -\lambda_{\Sigma} \langle \sigma\rangle \right)$ 
where $\mu_{\Sigma} \sim \mathcal{O} (M_{U})$. However depending upon the fine tuning involved in the model parameters, one can bring down fermion triplets much lower scale than the GUT scale. 
One can refer to Refs.~\cite{Chang:1983fu, Chang:1984uy, Chang:1984qr} for the details about the mass splitting among left-handed and right-handed scalar components which can be easily applicable to fermions also. We are however, assuming that in the parity breaking effects on the fermion triplets are smaller so that both can remain light with a tiny mass splitting. As discussed earlier, the origin of such effects can be speculated to be some unknown physics beyond the minimal LRSM or $SO(10)$ discussed here.

The one-loop renormalization group evolution (RGE) equation for gauge couplings is given by
\begin{equation}
\mu\,\frac{\partial g_{i}}{\partial \mu}=\frac{b_i}{16 \pi^2} g^{3}_{i},
\end{equation}
where the one-loop beta-coefficients $b_i$ are given by
\begin{eqnarray}
	&&b_i= - \frac{11}{3} \mathcal{C}_{2}(G) 
				 + \frac{2}{3} \,\sum_{R_f} T(R_f) \prod_{j \neq i} d_j(R_f) \nonumber \\
  &&\hspace*{2.5cm} + \frac{1}{3} \sum_{R_s} T(R_s) \prod_{j \neq i} d_j(R_s).
\label{oneloop_bi}
\end{eqnarray}

\begin{figure}[t]
\includegraphics[width=0.65\textwidth]{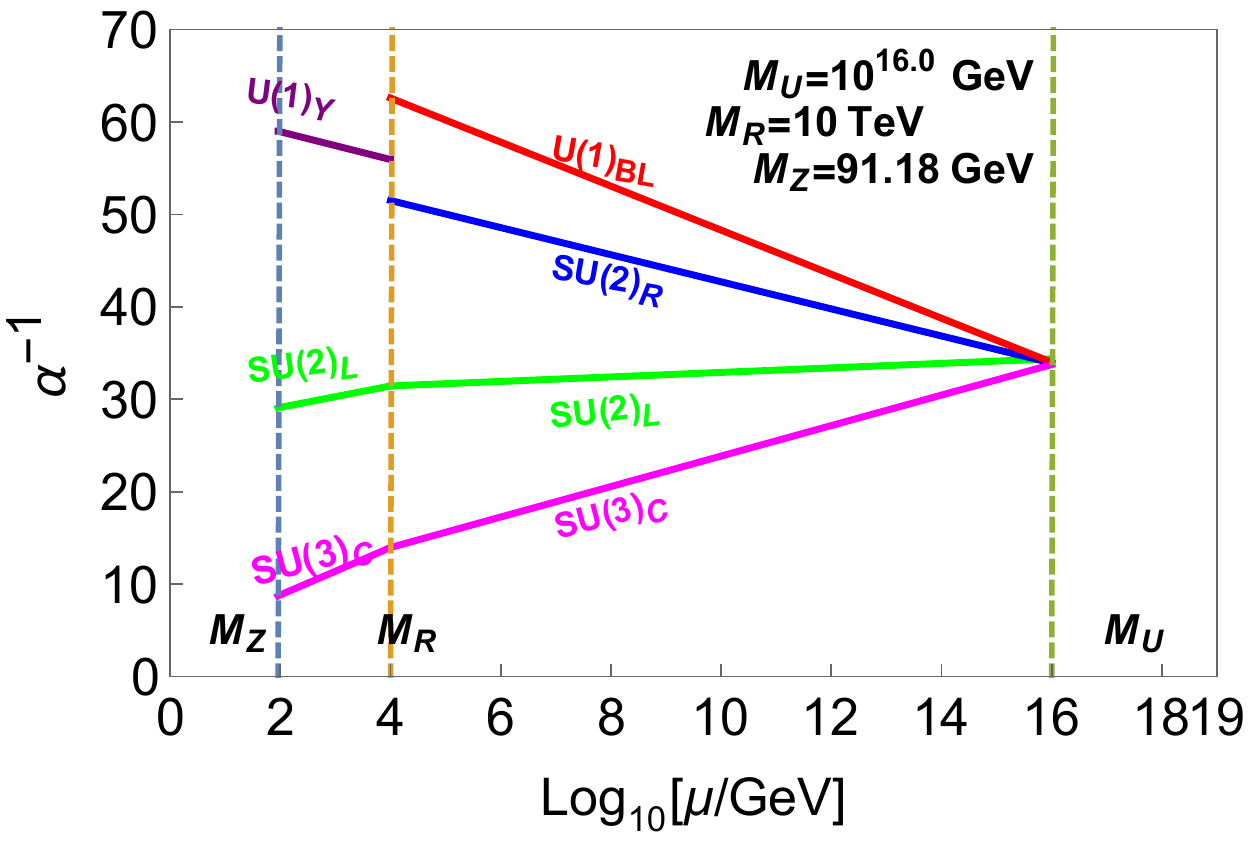}
\caption{Unification plot for three gauge couplings accommodating $W_R$ and fermion triplet 
around TeV scale where left-right symmetry breaks at $10~$TeV and $M_U\simeq 10^{16.0}$~GeV. 
We have added an extra color scalar field $\xi(1_L,3_R,-2/3_{BL},6_C)$ from $M_R$ to $M_U$ 
so as to get unification satisfying proton decay constraint.}
\label{fig:MvsMWR}
\end{figure}

Using the one loop RGE equations for gauge couplings and the derived values of beta 
coefficients, $b_i=\{-19/6, 41/10, -7\}$ from $M_Z-M_R$ and $b_i=\{-2/3, 4,13/2,-11/2\}$ from $M_R - M_U$, we 
show in figure \ref{fig:MvsMWR} that gauge coupling unification is possible and the predicted mass scales and 
gauge coupling ratio $g_R/g_L$ are given below
\begin{align}
M_U\simeq 10^{16.0}~\mbox{GeV}, \quad M_R\simeq \mbox{10 TeV},\quad \frac{g_R}{g_L}\simeq 0.78.
\end{align}

Since the left-right symmetry is broken at few TeV scale, we have extra neutral and charged gauge bosons 
around TeV scale which offers a rich collider studies \cite{Patra:2015bga} and in order to explain recent ATLAS and CMS 
anomalies including diphoton, diboson, dijet searches. The key feature of the model is the low scale $W_R$ and its 
discovery potential at LHC. The total cross-section for right-handed charged gauge boson $W_R$ production within LRSM 
for $M_{W_R}\simeq 2-3$~TeV and centre of mass energy $\sqrt{s}=8$~TeV 
is related to the mismatch between gauge couplings $g_R/g_L$ as following
\begin{align}
	\sigma(pp \to W_R) = 390\text{ fb} \cdot \left(\frac{g_R}{g_L}\right)^2.
\end{align}
It should be noted that the latest ATLAS data \cite{ATLAS16} do not show any significant excess in diboson channel. Also, the dark matter phenomenology 
discussed above suggests a heavier $W_R$ boson. A heavier $W_R$ boson can however, leave interesting signatures at the LHC, as discussed recently 
by the authors of \cite{manimala16}. As mentioned in \cite{manimala16}, a heavy $W_R$ boson, say of 5 TeV mass, can have a production cross section 
of around 2 fb at 13 TeV LHC whereas at 100 TeV proton proton future collider, even a much heavier $W_R$ boson with mass 25 TeV can be produced 
with a cross section of around 1 fb.

\section{Summary and Conclusion}
\label{sec6}
We have studied a minimal left-right symmetric model with fermion triplet dark matter candidates. The neutral component of these fermion triplets 
remain stable due to a remnant discrete symmetry $\mathcal{Z}_2\simeq (-1)^{B-L}$ to which the extended gauge symmetry of the model spontaneously 
breaks down to. The neutral components of both the triplets can simultaneously contribute to dark matter relic abundance, resulting in a multi-component 
dark matter scenario. The discrete left-right symmetry present in the model forces one to have the left and right dark matter masses equal at a high energy 
scale. However, spontaneous breaking of discrete parity at high scale can split the mass difference by an amount proportional to the scale of parity breaking. 
Assuming the effect of discrete parity breaking to be small on the triplet fermion masses due to the higher dimensional operators that arise by virtue of additional 
discrete symmetries, we consider a scenario where the neutral components of the fermion triplets are around the TeV corner but with a mass splitting of 3.55 keV. 
We introduce a scalar bitriplet into the model which assist the heavier dark matter decay at two loop level into the lighter one while emitting a photon of energy 
3.55 keV in order to explain the observations. We constrain the parameter space by keeping the mass difference between left and right dark matter to be 3.55 keV 
from the requirement of satisfying Planck data on dark matter relic abundance. We find that for right handed charged gauge boson mass 3, 4 TeV, we can satisfy 
relic abundance criteria for dark matter masses as low as a few hundred GeV's. Such low mass dark matter also keeps the perturbative calculation of dark matter 
relic abundance valid as the effects due to Sommerfeld's enhancement remains minimal. We also constrain the model parameters from the requirement of satisfying 
the constraint on decay width from X-ray data.

We then show how the scalar bitriplet which mediates two loop decay processes of heavier dark matter into the lighter one with a photon also assist in achieving 
gauge coupling unification at high energy scale while keeping the scale of left-right symmetry at TeV scale. We also comment how such a TeV scale LRSM can 
give rise to other interesting signatures at collider experiments. Such a TeV scale LRSM also has promising signatures at intensity frontier experiments like neutrinoless 
double beta decay, lepton flavour violation the detailed analysis of which can be found elsewhere.

\begin{acknowledgments}
DB acknowledges the support from IIT Guwahati start-up grant (reference number: xPHYSUGIITG01152xxDB001) and Associateship Programme of IUCAA, Pune.
\end{acknowledgments}

\appendix
\section{Three Body Decay of Heavier Dark Matter}
\label{appen1}
As we can see from the figure \ref{decayplot1} there are two loop functions. Let us call
the left hand side as $\mathcal{I}^{\rm bubble}$ and right hand side as 
$\mathcal{I}^{\rm vertex}$ which can be written as 
\begin{align*}
I^{\rm vertex}_{\mu \nu} &= \mu \int \frac{d^4l}{(2\pi)^4}\frac{(2l-k_3)^\nu (2l+k_2)^\mu}{(l^2-m^2_\psi)((l-k_3)^2-m^2_\psi)((l+k_2)^2-m^2_\psi)} \nonumber \\
I^{\rm bubble}_{\mu \nu} &= -\mu \int \frac{d^4l}{(2\pi)^4}\frac{g_{\mu \nu}}{(l^2-m^2_\psi)((l+ k_2 + k_3)^2-m^2_\psi)} \nonumber \\
I^{\rm vertex}_{\mu \nu} &= \frac{i\mu}{16\pi^2}\left[\mathcal{A}^{\rm vertex}g^{\mu \nu} + \mathcal{B}^{\rm vertex}k^\nu_2k^\mu_3 \right] \\
\mathcal{A}^{vertex} &= 3 + \frac{1}{\epsilon} + \ln \left[\frac{\mu^2}{r^2_\psi m^2_{\Sigma_R}}\right] + \frac{1}{x}\sqrt{x(x - 4r^2_{\psi})}\ln\left(\frac{2r^2_\psi - x + \sqrt{x(x - 4r^2_{\psi})}}{2r^2_{\psi}}\right) \\
&+ \frac{r^2_\psi}{x}\ln\left(\frac{2r^2_\psi - x + \sqrt{x(x - 4r^2_{\psi})}}{2r^2_{\psi}}\right)^2 \\
\mathcal{B}^{vertex} &= \frac{-2}{m^2_{\Sigma_R}r^2_\psi}\left(x + r^2_\psi\ln\left(\frac{2r^2_\psi - x + \sqrt{x(x - 4r^2_{\psi})}}{2r^2_{\psi}}\right)^2 \right) \\
I^{\rm bubble}_{\mu \nu} &= \frac{i\mu}{16\pi^2}\left[\mathcal{A}^{\rm bubble}g^{\mu \nu} + \mathcal{B}^{\rm Bubble}k^\nu_2k^\mu_3 \right] \\
\end{align*}
\begin{align*}
\mathcal{A}^{\rm bubble} &= -2 - \frac{1}{\epsilon} - \ln \left[\frac{\mu^2}{r^2_\psi m^2_{\Sigma_R}}\right] - \frac{1}{x}\sqrt{x(x - 4r^2_{\psi})}\ln\left(\frac{2r^2_\psi - x + \sqrt{x(x - 4r^2_{\psi})}}{2r^2_{\psi}}\right) \\
\mathcal{B}^{\rm bubble} &= 0 \\
I^{\rm scalar}_{\mu \nu} &= I^{\rm bubble}_{\mu \nu} + I^{\rm vertex}_{\mu \nu} \\
I^{\rm scalar}_{\mu \nu} &= \frac{i\mu}{16\pi^2}\left[\mathcal{A}^{\rm scalar}g^{\mu \nu} + \mathcal{B}^{\rm scalar}k^\nu_2k^\mu_3 \right] \\
\mathcal{A}^{\rm scalar}(x) &= \frac{1}{x}\left(x + r^2_\psi \ln\left[\frac{2r^2_\psi - x + \sqrt{x(x + 4r^2_\psi)}}{2r^2_\psi}\right]^2\right)\\
&\simeq \frac{-x}{12 r^2_\psi} + \mathcal{O}(x^{3/2}) \\
\mathcal{B}^{\rm scalar}(x) &=  -\frac{2}{x^2}\left(x + r^2_\psi \ln \left[\frac{2r^2_\psi - x + \sqrt{x(x + 4r^2_\psi)}}{2r^2_\psi}\right]^2\right)\\
&\simeq \frac{1}{6 m^2_{\Sigma_R}r^2_\psi} + \frac{x}{45 m^2_{\Sigma_R}r^4_\psi} + \mathcal{O}(x^{3/2}) \\
r_\psi &= \frac{m_\psi}{m_{\Sigma_R} } \quad x = \frac{2k_2.k_3}{m_{\Sigma_R}}
\end{align*}
Similarly for the loop correction coming from the top quark loop 
due to the scalar mixing to the SM Higgs is as follows
\begin{align}
I^{\rm fermion}_{\mu \nu} &= -\theta \frac{m_t^2}{v}\int \frac{d^4l}{(2\pi)^4}\frac{16l^\mu l^\nu - 4g^{\mu \nu}(l^2 + k_2.k_3 - m^2) + 4k_2^{\nu}k_3^{\nu}}{(l^2-m^2_\psi)((l-k_3)^2-m^2_\psi)((l+k_2)^2-m^2_\psi)} \nonumber \\
I^{\rm fermion}_{\mu \nu} &= -\theta \frac{m_t^2}{v}\frac{i}{16\pi^2}\left[\mathcal{A}^{\rm fermion}g^{\mu \nu} + \mathcal{B}^{\rm fermion}k^\nu_2k^\mu_3 \right] \nonumber \\
\mathcal{A}^{\rm fermion}(x) &= \frac{1}{x}\left(4x + (4r^2_t - x)\ln\left[\frac{2r_t^2 - x + \sqrt{x(x- 4 r_t^2)}}{2r^2_t}\right]^2\right)\nonumber \\
\mathcal{B}^{\rm fermion}(x) &= -\frac{2}{x}\mathcal{A}^{\rm fermion}
\end{align}
And the loop contribution hence is coming out to be
\begin{align*}
\mathcal{I}(x) &= \frac{1}{256\pi^4}\left[4|\mathcal{A}(x)|^2 + x\Re(\mathcal{A}^*(x)\mathcal{B}(x))\right] \end{align*}
where $\mathcal{A} = \mathcal{A}^{\rm fermion} + \mathcal{A}^{\rm scalar}$ and $\mathcal{B} = \mathcal{B}^{\rm fermion} + \mathcal{B}^{\rm scalar}$. Now, the Decay width is coming out to be
\begin{align*}
\Gamma_{\Sigma^0_R \rightarrow \Sigma^0_L \gamma \gamma} &= \frac{Y^2_{\psi}e^4 }{256\pi^3 m_{\Sigma^0_R}} \int^{1 + r^2_{\Sigma^0_L}}_{2r_{\Sigma^0_L}}  \frac{2(z-2r^2_{\Sigma^0_L})\sqrt{z^2 - 4 r^2_{\Sigma^0_L}} }{(1 + r^2_{\Sigma^0_L} -z - r^2_\psi)^2}|\mathcal{I}(1 + r^2_{\Sigma^0_L} - z)|^2 dz\\
&\simeq \frac{Y^2_{\psi}e^4}{256\pi^3 m^2_{\Sigma^0_R}} \left[\frac{\Delta k^9 \left(8m^2_t m^2_\psi\theta + m^2_t v\mu \right)^2}{90720\pi^4 m_t^4 v^2 m^8_{\psi}}\right]
\end{align*}
where $x = 2p.k_1/m^2_{\Sigma^0_R}$ and $r_{\Sigma^0_L} = \frac{m_{\Sigma^0_L}}{m_{\Sigma^0_R}},  \Delta k = 3.55$ keV and $\theta$ is the mixing parameter between bitriplet scalar and SM Higgs.

Taking the part of the amplitude $\mathcal{I}_{\mu \nu}$, we also check if it satisfies the Ward-Takahashi Identity. For that we need to see if $k_2^\mu\mathcal{I}_{\mu \nu} = k_3^\nu\mathcal{I}_{\mu \nu} = 0$. We can show that
\begin{align*}
k_2^\mu\mathcal{I}_{\mu \nu} &= \frac{i}{16\pi^2}\left[k_{2\nu}\mathcal{A}(x) + k_2.k_3 k_{2\nu}\mathcal{B}(x)\right] \nonumber \\
&= \frac{ik_{2\nu}}{16\pi^2}\left[\mathcal{A}(x) + \frac{x}{2}\mathcal{B}(x)\right] = 0 \nonumber 
\end{align*}
as one can see from the above expressions for $A(x), B(x)$ that $\mathcal{A}(x) = \frac{-x}{2}\mathcal{B}(x)$. Similarly, one can see $k_3^\nu\mathcal{I}_{\mu \nu} = 0$ also.

\bibliographystyle{apsrev}

\end{document}